\documentclass[pra,onecolumn,superscriptaddress,amsmath,amssymb,aps,floatfix]{revtex4-2}
\usepackage{graphicx}
\usepackage{dcolumn}
\usepackage{bm}
\usepackage{amsmath,amssymb}
\usepackage{graphics}
\usepackage{amsfonts}
\usepackage{epsfig}
\usepackage{float}
\usepackage{hyperref}
\usepackage{adjustbox}
\usepackage{rotating}
\usepackage{float}
\usepackage{dblfloatfix}

\usepackage{blindtext}
\usepackage{multirow}
\usepackage{graphicx}
\usepackage{dcolumn}
\usepackage{bbold}
\usepackage{bm}
\usepackage{amsmath}
\usepackage{import}
\usepackage{physics}
\usepackage{verbatim}
\usepackage{listings}
\usepackage{xcolor}
\usepackage{xr-hyper}
\usepackage{hyperref}
\usepackage{cleveref}
\usepackage{newfloat}
\usepackage{caption}
\usepackage{subcaption}
\DeclareFloatingEnvironment[name={Fig S},fileext=lof]{suppfigure}
\begin{document}
\title{Yu-Shiba-Rusinov bound states boost surface-induced odd-frequency superconductivity}
\author{Subhajit Pal}
\author{Colin Benjamin} \email{colin.nano@gmail.com}
\affiliation{School of Physical Sciences, National Institute of Science Education \& Research, Jatni-752050, India.}
\affiliation{Homi Bhabha National Institute, Training School Complex, AnushaktiNagar, Mumbai, 400094, India.}
\begin{abstract}
We predict that the occurence of zero energy Yu-Shiba-Rusinov(YSR) bound states in two different setups, metal-spin flipper-metal-s-wave superconductor ($N_{1}-sf-N_{2}-S$) and superconductor-metal-spin flipper-metal-superconductor ($S-N_{1}-sf-N_{2}-S$) junctions, can generate multi-fold enhancement of surface-induced odd-frequency superconductivity. On the other hand, in the absence of these bound states, even-frequency superconductivity dominates. Specifically, in a $S-N_{1}-sf-N_{2}-S$ Josephson junction, the emergence of zero energy YSR bound states leads to a $0-\pi$ junction transition and surface odd-frequency superconductivity dominance. Notably, odd-frequency superconductivity vanishes in the absence of YSR-bound states. Interestingly, the equal spin-triplet pairing is the dominant component in the surface induced odd-frequency superconductivity in both setups, which could have important implications for superconducting spintronics. Overall, our findings may help to detect the presence of YSR-bound states through the observation of surface induced odd-frequency superconductivity and contribute to a better understanding of their relationship.
\end{abstract}
\maketitle
\section{Introduction} Odd-frequency superconductivity is a fascinating phenomenon that has been extensively studied in recent years. One of the key features of odd-frequency superconductivity is the sign change in Cooper pair wavefunctions, which is a consequence of the exchange of time coordinates between the two electrons in the pair. This is in contrast to even-frequency superconductivity, where electron pairing occurs at equal times. Even-frequency Cooper pairs can be further classified into even-frequency spin-singlet (Even-SS) and even-frequency spin-triplet (Even-ST) pairing states. Examples of Even-SS pairing are $s$ and $d$ wave pairing, while the example of Even-ST pairing is $p$ wave pairing\cite{sig}. In contrast, odd-frequency Cooper pairing which was initially discovered in $^3$He\cite{bere}, can be either in an odd-frequency spin-singlet (Odd-SS) state or an odd-frequency spin-triplet (Odd-ST) state. A spin-triplet state, regardless of whether its even- or odd-frequency, can be either mixed spin-triplet (MST), i.e., $|\uparrow\downarrow\rangle + |\downarrow\uparrow\rangle$ or equal spin-triplet (EST), i.e., $|\uparrow\uparrow\rangle$, $|\downarrow\downarrow\rangle$. The spin-singlet (SS) state is of one type, i.e., $|\uparrow\downarrow\rangle - |\downarrow\uparrow\rangle$. {Odd-SS pairing can be induced in metal-superconductor junction due to the breaking of spatial parity at metal-superconductor interfaces\cite{tanaka-2007}. Further, odd-MST pairing has been predicted to arise in a metal-superconductor junction with spin-active interface\cite{linn}.}
{Theoretically, odd-EST pairing was first predicted to occur in diffusive ferromagnet-superconductor junction with inhomogeneous magnetization\cite{fsb1,fsb2}. Odd-frequency spin-triplet superconductivity is also induced in a multilayered superconductor-ferromagnet structures with a  noncollinear alignment of the magnetizations\cite{fsb3,fsb4}. Further, in a superconductor/half-metal/superconductor junction, spin rotation induces odd-frequency spin-triplet pairing in the superconductor which results in an indirect Josephson effect in presence of surface spin flip scattering\cite{mes}.} One interesting aspect of odd-frequency superconductivity\cite{jli,ctr,tana} is its potential applications in spintronics. For example, Odd-EST pairing induced by magnetic impurities at a metal-superconductor interface has been shown to generate a spin current\cite{odd}. Moreover, magnetic impurities can also induce YSR bound states.
It was first noticed by Yu, Shiba, and Rusinov in the late 1960s and is now called the Yu-Shiba-Rusinov (YSR) state\cite{Yu,Shib,Rusi}. The interaction of Andreev reflected electrons or holes with impurity spin gives rise to these YSR states, see Refs.~\cite{ysr,pers,costa}. YSR bound states have been observed experimentally in various superconducting materials\cite{yaz,frank,Ruby,hji,ran,lcor,hat,sen}. {For example, in Ref.~\cite{RUBY}, the authors study the microscopic tunneling processes between a scanning tunneling microscope (STM) tip and a YSR bound state of a single magnetic impurity which is doped on the surface of a superconductor. They observe different subgap spectra by changing the distance between the STM tip and the sample. Further, in Ref.~\cite{cor}, YSR bound states are observed in superconducting graphene without the need of any magnetic impurity. The authors induce superconductivity in graphene and exploit graphene grain boundaries as a source of local magnetic moments to realize YSR bound states in graphene.}

In this work, our primary motivation is to understand if there exists any link between {surface} odd-frequency pairing and YSR-bound states. We notice that for parameter regimes wherein zero energy YSR bound states appear, {surface} odd-frequency pairing dominates over even-frequency pairing. However, for parameter regimes wherein zero energy YSR bound states are absent, even-frequency pairing dominates over {surface} odd-frequency pairing. An enhancement of {surface} odd-frequency pairing due to YSR-bound states implies a nontrivial link between these remarkable effects. Furthermore, the odd-frequency superconductivity generated in our setups is equal spin-triplet pairing due to interface Andreev scattering. Odd-frequency mixed spin-triplet (Odd-MST) pairing\cite{suzu,perrin} has also been reported in the vicinity of magnetic impurity. {In Ref.~\cite{mats}, it is shown that odd-MST pairing emerges at the edge of a $d$-wave superconductor due to the surface Andreev bound states.}

EST pairing, unlike MST pairing, has significant implications for superconducting spintronics, as spin is much more resilient to dephasing. In this paper, when zero energy YSR bound states appear, Odd-EST pairing dominates Even-EST pairing, while Odd-MST pairing and Even-MST pairings vanish. EST pairing supports dissipationless pure spin current, which has applications in superconducting spintronics\cite{esc,lin}. Further, via inducing Odd-EST pairing, one can effectively tune an even $\omega$ $s$-wave superconductor to an odd $\omega$ $p$-wave superconductor.

This paper is organized into five sections, including an appendix. The first section presents the two setups and provides the theoretical background for the study, {which are necessary} for calculating various quantities such as YSR bound states, Josephson current, and Green's functions. The procedure for calculating YSR bound states, Josephson current, and induced pairing amplitude from retarded Green's functions is also discussed in this section. The second section presents the results of the study and explains the relationship between odd-frequency pairing and YSR bound states in metal-spin flipper-metal-superconductor junction and superconductor-metal-spin flipper-metal-superconductor junction. The third section provides an analysis of the results via a table. The fourth section concludes the paper with an experimental realization. {Finally, the Appendix provides the wavefunctions of our two systems, the effect of interface transparency and finite temperature on superconducting pairing, and the explicit form of expressions for anomalous Green's functions.}
\section{Theory}
\subsection{YSR bound states in metal-Superconductor and Superconductor-metal-Superconductor junctions with spin flipper}
\subsubsection{Hamiltonian}
We choose two setups: (a) 1D metal (N)-metal (N)-$s$-wave superconductor (S) junction with a spin-flipper (sf) between the two metals as depicted in Fig.~1(a) and (b) two metals with an embedded spin-flipper between two $s$-wave superconductors as shown in Fig.~1(b). We use the BTK approach\cite{BTK} to solve the problem. Spin-flipper is a $\delta$-like magnetic impurity. The Hamiltonian for spin-flipper is\cite{AJP,Liu,Maru,FC,ysr}:
\begin{equation}
H_{\mbox{spin-flipper}}=-\mathcal{J}_{0}\delta(x)\vec{s}\cdot\vec{S}.
\label{flipper}
\end{equation}
{Our model spin flipper is different from a Kondo-like magnetic impurity\cite{suzu}, which has its own dynamics. For Kondo-like magnetic impurities randomly distributed in an $s$-wave superconductor, Odd-MST pairing is induced\cite{flv}. However, in our case the spin flipper generates Odd-EST pairing with vanishing Odd-MST pairing.}
The model Hamiltonian in Bogoliubov-de Gennes (BdG) formalism for metal-spin-flipper-metal-superconductor (N$_{1}$-sf-N$_{2}$-S) junction as shown in Fig.~1(a) {is a $4\times4$ matrix and given as,}
\begin{equation}
H_{BdG}^{\mbox{N$_{1}$-sf-N$_{2}$-S}}(x)=
\begin{pmatrix}
H_{M}\hat{I} & i\Delta \theta(x)\hat{\sigma}_{y} \\
-i\Delta^{*}\theta(x)\hat{\sigma}_{y} & -H_{M}\hat{I}
\end{pmatrix},
\label{ham}
\end{equation}
where $H_{M}={-\frac{\hbar^2}{2m^{*}}\frac{\partial^2}{\partial x^2}}+V\delta(x)-\mathcal{J}_{0}\delta(x+a)\vec s.\vec S -E_{F}$, $\theta(x)$ is the Heaviside step function, $\Delta$ is the gap parameter for $s$-wave superconductor. {$-\frac{\hbar^2}{2m^{*}}\frac{\partial^2}{\partial x^2}$ is electron's kinetic energy operator with effective mass $m^{*}$}, $V$ is strength of the $\delta$-type potential barrier at the interface between metal and superconductor, the third term represents the exchange coupling of strength $\mathcal{J}_{0}$ between electron's spin ($\vec{s}$) and spin ($\vec{S}$) of spin flipper, $\hat{I}$ is an unit matrix, $\hat{\sigma}$ represents the Pauli matrices and $E_F$ denotes the Fermi energy. Spin flipper's spin can be in all possible states. For example if the spin flipper's spin $S=1/2$, then spin flipper's spin can have two possible states, $m'=1/2$ and $m=-1/2$. Similarly, if $S=3/2$, then spin flipper's spin can have four possible states, $m'=3/2,1/2,-1/2,-3/2$. We calculate different measurable quantities like differential charge conductance/Josephson current/superconducting pairing magnitude for each of the $2S+1$ possible values of $m'$ for spin flipper's spin $S$ and finally take an average over all $m'$ values. {When an spin up electron scatters from the spin flipper (with spin $S=1/2$), a product state ($\ket{\uparrow}_{e}\times\ket{\uparrow}_{SF}$) will form, if spin flipper's spin is in up state ($m'=1/2$), while an entangled state ($\ket{\downarrow}_{e}\otimes\ket{\uparrow}_{SF}+\ket{\uparrow}_{e}\otimes\ket{\downarrow}_{SF}$) will form after scattering if spin flipper's spin is in spin down state ($m'=-1/2$). For the next scattering event, at spin flipper, if spin down electron scatters from the spin flipper, it sees spin flipper either in spin up state or spin down state. In case spin flipper is in spin up state then spin flip occurs and an entangled state is formed. Thus, measurable quantities like differential charge conductance/Josephson current/pairing magnitude will be the average of the two processes: spin flip \& no flip. A similar thing happens for $S=3/2$ where there are four possible states, $m'=3/2, 1/2,-1/2,-3/2$ leading to one no flip process $S=m'=3/2$ and other three flip processes $S=3/2$ and $m'=1/2,-1/2,-3/2$. Also for other higher spin states of spin flipper's spin $S$, one calculates these measurable quantities by averaging over all possible values of $m'$.} In this paper, the dimensionless parameter $\mathcal{J}=\frac{m^{*}\mathcal{J}_{0}}{k_{F}}$ is used as a measure of strength of exchange coupling\cite{AJP} and $Z=\frac{m^{*}V}{\hbar^2 k_{F}}$ as a measure of interface transparency\cite{BTK}.

The model Hamiltonian in BdG formalism for superconductor-metal-spin-flipper-metal-superconductor (S-N$_{1}$-sf-N$_{2}$-S) junction as shown in Fig.~1(b) {is a $4\times4$ matrix and given as,}
\begin{equation}
H_{BdG}^{\mbox{S-N$_{1}$-sf-N$_{2}$-S}}(x)=
\begin{pmatrix}
H_{J}\hat{I} & i\Delta_{J} \hat{\sigma}_{y} \\
-i\Delta_{J}^{*} \hat{\sigma}_{y} & -H_{J}\hat{I}
\end{pmatrix},
\label{hamm}
\end{equation}
where $H_{J}=p^2/2m^{*}+V[\delta(x+a/2)+\delta(x-a/2)]-J_{0}\delta(x)\vec s.\vec S-E_{F}$. The superconducting gap $\Delta_{J}$ is of the form $\Delta_{J}=\Delta[e^{i\varphi_{L}}\Theta(-x-a/2)+e^{i\varphi_{R}}\Theta(x-a/2)]$. $\varphi_{L}$ and $\varphi_{R}$ are the superconducting phases for left and right superconductors respectively.
\begin{figure}[h]
\centering{\includegraphics[width=0.99\textwidth]{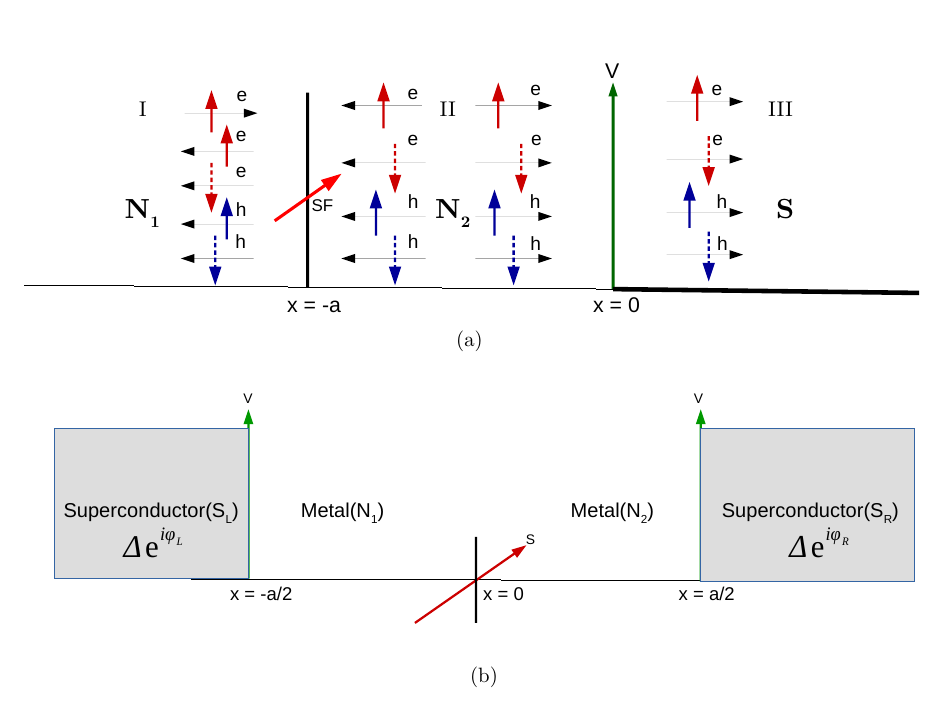}}
\caption{\small \sl (a) N$_{1}$N$_{2}$S junction with spin flipper at N$_{1}$N$_{2}$ interface and a $\delta$-like potential barrier at N$_{2}$S interface. The scattering of a spin up electron incident is shown, (b) Josephson junction composed of two metals and a spin flipper with spin $S$ and magnetic moment $m'$ at $x=0$ embedded between two $s$-wave superconductors. {In (a) and (b) spin flipper's spin is in arbitrary direction having components in $x$, $y$ and $z$ directions.}}
\end{figure}
{After diagonalizing the Hamiltonians (Eqs.~\eqref{ham},\eqref{hamm}), the wavefunctions in different domains of metal-spin flipper-metal-superconductor junction and superconductor-metal-spin flipper-metal-superconductor junction for various kinds of scattering processes are obtained. The wavefunctions in different regions of our setups are mentioned in Appendix.}
\subsubsection{Boundary conditions for metal-spin flipper-metal-superconductor junction}
From the boundary condition, we get the amplitudes in the scattering states. At $x=-a$, the boundary conditions are-
\begin{equation}
\label{bc1}
\varphi_{i}|_{x<-a}=\varphi_{i}|_{-a<x<0}\,\,\, \mbox{and},\,\,\,
\frac{d\varphi_{i}|_{-a<x<0}}{dx}-\frac{d\varphi_{i}|_{x<-a}}{dx}=-\frac{2m^{*}\mathcal{J}_{0}\vec{s}.\vec{S}}{\hbar^2} \varphi_{i}|_{x=-a},
\end{equation}
where $\vec{s}.\vec{S} { =s^{x}S^{x}+s^{y}S^{y}+s^{z}S^{z}}=s^{z}S^{z}+\frac{(s^{+}S^{-}+s^{-}S^{+})}{2}$, $s^{\pm}=s_{x}\pm is_{y}$ and $S^{\pm}=S_{x}\pm iS_{y}$ are the exchange operator in Hamiltonian for spin flipper\cite{AJP}, spin raising \& lowering operators for electron and spin raising \& lowering operators for spin flipper respectively. The boundary conditions at $x=0$ are-
\begin{equation}
\varphi_{i}|_{-a<x<0}=\varphi_{i}|_{x>0}
\mbox{ and, }\,\,\frac{d\varphi_{i}|_{x>0}}{dx}-\frac{d\varphi_{i}|_{-a<x<0}}{dx}=\frac{2m^{*}V}{\hbar^2}\varphi_{i}|_{x=0}.
\label{bc4}
\end{equation}
{Operation of $\vec{s}.\vec{S}$ for wave-function associating up spin electron is
\begin{equation}
\label{EUU}
\vec s.\vec S\chi_{1}^{N}\phi_{m'}^{S}=s_{z}S_{z}\chi_{1}^{N}\phi_{m'}^{S}+\frac{1}{2}s^{-}S^{+}
              \chi_{1}^{N}\phi_{m'}^{S}+\frac{1}{2}s^{+}S^{-}\chi_{1}^{N}\phi_{m'}^{S},
\end{equation}
Now, $s^{+}\chi_{1}^{N}=0$, because $s^{+}$ is the electron's spin raising operator and no higher spin states are possible for an electron with spin $1/2$ than up and thus the third term in Eq.~\eqref{EUU} becomes zero, while the spin lowering operator $s^{-}\chi_{1}^{N}=\hbar\chi_{2}^{N}$ ($\chi_{2}^{N}=(0\,1\, 0\, 0)^{T})$ gives the spin down state $\chi_{2}^{N}$ of electron. In addition, for up spin electron $s_{z}\chi_{1}^{N}=\frac{\hbar}{2}\chi_{1}^{N}$ and for spin flipper $S^{z}\phi_{m'}^{S}=\hbar m'\phi_{m'}^{S}$. The spin raising operator operating on spin flipper provide: $S^{+}\phi_{m'}^{S}=\hbar\sqrt{(S-m')(S+m'+1)}\phi_{m'+1}^{S}=\hbar\mathcal{F}\phi_{m'+1}^{S}$. Thus from Eq.~\eqref{EUU} we get,}
\begin{equation}
\vec{s}.\vec{S}\chi_{1}^{N}\phi_{m'}^{S}=\frac{\hbar^2m'}{2}\chi_{1}^{N}\phi_{m'}^{S}+\frac{\hbar^2\mathcal{F}}{2}\chi_{2}^{N}\phi_{m'+1}^{S}.
\label{eu}
\end{equation}
Similarly, for wave-function associating down spin electron, action of $\vec{s}.\vec{S}$ is
\begin{equation}
\label{ed}
\vec{s}.\vec{S} \chi_{2}^{N}\phi_{m'}^{S}=-\frac{\hbar^2m'}{2}\chi_{2}^{N}\phi_{m'}^{S}+\frac{\hbar^2\mathcal{F'}}{2}\chi_{1}^{N}\phi_{m'-1}^{S}.
\end{equation}
Further, for wave-function associating up spin hole, action of $\vec{s}.\vec{S}$ is
\begin{equation}
\label{hu}
\vec{s}.\vec{S}\chi_{3}^{N}\phi_{m'}^{S}=-\frac{\hbar^2m'}{2}\chi_{3}^{N}\phi_{m'}^{S}+\frac{\hbar^2\mathcal{F'}}{2}\chi_{4}^{N}\phi_{m'-1}^{S},
\end{equation}
and finally, for wave-function associating down spin hole, action of $\vec{s}.\vec{S}$ is
\begin{equation}
\label{hd}
\vec{s}.\vec{S}\chi_{4}^{N}\phi_{m'}^{S}=\frac{\hbar^2m'}{2}\chi_{4}^{N}\phi_{m'}^{S}+\frac{\hbar^2\mathcal{F}}{2}\chi_{3}^{N}\phi_{m'+1}^{S}.
\end{equation}
In Eqs.~\eqref{eu}-\eqref{hd}, $\mathcal{F}=\sqrt{(S-m')(S+m'+1)}$ and $\mathcal{F'}=\sqrt{(S+m')(S-m'+1)}$ are the flip probabilities when spin up(or, down) electron(or, hole) is incident and, when spin down (or, up) electron(or, hole) is incident respectively. After using above equations and solving boundary conditions, we will get $16$ equations for each kind of incident process, see Eq.~\eqref{wav}. Different scattering amplitudes $\it{a}_{ij}$, $\it{b}_{ij}$, $\it{c}_{ij}$, $\it{d}_{ij}$, $\it{e}_{ij}$, $\it{f}_{ij}$, $\it{g}_{ij}$, $\it{h}_{ij}$ for each kind of scattering process are obtained from these $16$ equations. We use these scattering amplitudes to compute energy bound states and retarded Green's function for the setup shown in Fig.~1(a). Even-/odd-frequency spin-singlet and spin-triplet pairings are found from retarded Green's function.
\subsubsection{Boundary conditions for superconductor-metal-spin flipper-metal-superconductor junction}
At $x=-a/2$, the boundary condition is-
\begin{eqnarray}
\label{bc5}
{}&\varphi_{i}(x<-a/2)=\varphi_{i}(-a/2<x<0),\\
&\mbox { and, }\frac{d\varphi_{i}(-a/2<x<0)}{dx}-\frac{d\varphi_{i}(x<-a/2)}{dx}=\frac{2m^{\star}V}{\hbar^2} \varphi_{i}(x=-a/2).
\end{eqnarray}
Similarly, at $x=0$, the boundary condition is-
\begin{eqnarray}
{}&\varphi_{i}(-a/2<x<0)=\varphi_{i}(0<x<a/2),\\
&\mbox { and, }\frac{d\varphi_{i}(0<x<a/2)}{dx}-\frac{d\varphi_{i}(-a/2<x<0)}{dx}=-\frac{2m^{\star}J_{0}\vec s.\vec S}{\hbar^2} \varphi_{i}(x=0).
\end{eqnarray}
Finally, at $x=a/2$, the boundary condition is-
\begin{eqnarray}
{}&\varphi_{i}(0<x<a/2)=\varphi_{i}(x>a/2),\\
&\mbox { and, }\frac{d\varphi_{i}(x>a/2)}{dx}-\frac{d\varphi_{i}(0<x<a/2)}{dx}=\frac{2m^{\star}V}{\hbar^2} \varphi_{i}(x=a/2).
\label{bc55}
\end{eqnarray}
Using Eqs.~(\ref{eu}-\ref{hd}) and solving boundary conditions at $x=-a/2$, $x=0$ and $x=a/2$, 24 equations for each kind of incident process as mentioned in Eq.~\eqref{wavv} are obtained. Using these 24 equations, the various scattering amplitudes $a'_{mn}$, $b'_{mn}$, $c'_{mn}$, $d'_{mn}$, $e'_{mn}$, $f'_{mn}$, $g'_{mn}$, $h'_{mn}$, $i'_{mn}$, $j'_{mn}$, $k'_{mn}$, $l'_{mn}$ for each kind of scattering process can be obtained.
\subsubsection{YSR bound states in metal-spin flipper-metal-superconductor junction}
To calculate YSR bound states for this case, we first compute differential charge conductance using the well-established definitions as\cite{cheng,kash}
{
\begin{equation}
\mathcal{G}_{c}=\mathcal{G}_{N}(2+\mathcal{A}_{11}+\mathcal{A}_{12}-\mathcal{B}_{11}-\mathcal{B}_{12}+\mathcal{A}_{21}+\mathcal{A}_{22}-\mathcal{B}_{21}-\mathcal{B}_{22}),
\label{cond}
\end{equation}}
where $\mathcal{G}_{N}=e^2/h$ is the charge conductance when a metallic region replaces the superconducting region in our model with $\Delta=0$. {$\mathcal{A}_{1(2)1}(=\frac{q_{h}}{q_{e}}|\it{a_{1(2)1}}|^2)$} is the Andreev reflection probability when a spin-up {(down)} electron is reflected as a spin-up hole, {$\mathcal{A}_{1(2)2}(=\frac{q_{h}}{q_{e}}|\it{a_{1(2)2}}|^2)$} is the Andreev reflection probability when a spin-up {(down)} electron is reflected back as a spin down hole, {$\mathcal{B}_{1(2)1}(=|\it{b_{1(2)1}}|^2)$} is the normal reflection probability when a spin-up {(down)} electron is reflected as a spin-up electron and, finally {$\mathcal{B}_{1(2)2}(=|\it{b_{1(2)2}}|^2)$} is the normal reflection probability when a spin-up {(down)} electron is reflected as a spin-down electron. From complex poles of differential charge conductance $\mathcal{G}_{c}$ in Eq.~\eqref{cond}, we can get energy bound states $E^{\pm}$. The real part of the poles is the energy where YSR peaks appear, while the imaginary part of the poles denotes the width of the peaks.
\subsubsection{YSR bound states in superconductor-metal-spin-flipper-metal-superconductor junction}
An electron in the metallic region is incident at the NS interface with an energy below the superconducting gap that cannot penetrate the superconductor. However, at the NS interface, Andreev's reflection may happen, in which a hole with opposite momentum is reflected into normal metal, and a Cooper pair is generated in the superconductor. The same effect is also present with an identical electron/hole combination at the SN interface. Therefore a bound state is formed between the two superconductors, an Andreev bound state. From Andreev bound states, one can calculate Free energy and Josephson current through the junction. To calculate energy bound states in superconductor-metal-spin flipper-metal-superconductor junction we ignore the contribution from incoming quasiparticle\cite{linder,annu} and put the wavefunction into the boundary conditions. We obtain $24$ equations for the scattering amplitudes. Eliminating the scattering amplitudes of the two metals by the scattering amplitudes in the left and right superconductors we get $8$ equations,
\begin{equation}
Lz=0,
\end{equation}
where $z$ is a $8\times1$ column matrix and given by $z=[b'_{11},b'_{12},a'_{11},a'_{12},k'_{11},k'_{12},l'_{11},l'_{12}]^{T}$ and $L$ being a $8\times8$ matrix. For nontrivial solution of this system, the determinant of $L$ is zero (det $L=0$) and we get energy bound states $E_{k}(\varphi)$ ($k=\{1,...,4\}$) as a function of phase difference $\varphi$ between right and left superconductor. We see that $E_{k}(\varphi)=E_{\sigma}^{\pm}(\varphi)=\pm E_{\sigma}(\varphi), (\sigma=\uparrow,\downarrow)$. Our work considers the short junction limit ($a\ll\xi$, where $\xi$ is the superconducting coherence length). Thus total Josephson current is the same as the Josephson bound state current.
The Josephson current can be calculated from Andreev energy bound states\cite{goluu},
\begin{equation}
I=\frac{2e}{\hbar}\sum_{k}f(E_{k})\frac{dE_{k}}{d\varphi}=-\frac{2e}{\hbar}\sum_{\sigma}\tanh\Big(\frac{\beta E_{\sigma}}{2}\Big)\frac{dE_{\sigma}}{d\varphi}.
\end{equation}
\subsection{Retarded and Advanced Green's functions in presence of YSR bound states}
The main aim of our work is to check whether the presence of YSR-bound states has any bearing on the odd $\omega$ pairing generated. To this end we construct retarded Green's function $G^{r}(x,x',\omega)$ for our setups shown in Figs.~1(a) \& (b) from the scattering processes at the interface\cite{mcm}.
We follow Refs.~\onlinecite{cayy, amb} and the retarded Green's function is given as\cite{mcm,furu,kta}-
\begin{equation}
\label{RGF}
\begin{split}
G^{r}(x,x',\omega)=
\begin{cases}
\varphi_{1}(x)[\alpha_{11}\tilde{\varphi}_{5}^{T}(x')+\alpha_{12}\tilde{\varphi}_{6}^{T}(x')+\alpha_{13}\tilde{\varphi}_{7}^{T}(x')+\alpha_{14}\tilde{\varphi}_{8}^{T}(x')]\\
+
\varphi_{2}(x)[\alpha_{21}\tilde{\varphi}_{5}^{T}(x')+\alpha_{22}\tilde{\varphi}_{6}^{T}(x')+\alpha_{23}\tilde{\varphi}_{7}^{T}(x')+\alpha_{24}\tilde{\varphi}_{8}^{T}(x')]\\
+
\varphi_{3}(x)[\alpha_{31}\tilde{\varphi}_{5}^{T}(x')+\alpha_{32}\tilde{\varphi}_{6}^{T}(x')+\alpha_{33}\tilde{\varphi}_{7}^{T}(x')+\alpha_{34}\tilde{\varphi}_{8}^{T}(x')]\\
+
\varphi_{4}(x)[\alpha_{41}\tilde{\varphi}_{5}^{T}(x')+\alpha_{42}\tilde{\varphi}_{6}^{T}(x')+\alpha_{43}\tilde{\varphi}_{7}^{T}(x')+\alpha_{44}\tilde{\varphi}_{8}^{T}(x')]
\,,\quad x>x'&\\
\varphi_{5}(x)[\beta_{11}\tilde{\varphi}_{1}^{T}(x')+\beta_{12}\tilde{\varphi}_{2}^{T}(x')+\beta_{13}\tilde{\varphi}_{3}^{T}(x')+\beta_{14}\tilde{\varphi}_{4}^{T}(x')]\\
+\varphi_{6}(x)[\beta_{21}\tilde{\varphi}_{1}^{T}(x')+\beta_{22}\tilde{\varphi}_{2}^{T}(x')+\beta_{23}\tilde{\varphi}_{3}^{T}(x')+\beta_{24}\tilde{\varphi}_{4}^{T}(x')]\\
+\varphi_{7}(x)[\beta_{31}\tilde{\varphi}_{1}^{T}(x')+\beta_{32}\tilde{\varphi}_{2}^{T}(x')+\beta_{33}\tilde{\varphi}_{3}^{T}(x')+\beta_{34}\tilde{\varphi}_{4}^{T}(x')]\\
+\varphi_{8}(x)[\beta_{41}\tilde{\varphi}_{1}^{T}(x')+\beta_{42}\tilde{\varphi}_{2}^{T}(x')+\beta_{43}\tilde{\varphi}_{3}^{T}(x')+\beta_{44}\tilde{\varphi}_{4}^{T}(x')]\,, \quad x<x'&
\end{cases}
\end{split}
\end{equation}
In Eq.~\eqref{RGF}, $\alpha_{ij}$ and $\beta_{mn}$ are computed from the continuity of the Green's function
\begin{equation}
[\omega-H_{BdG}(x)]G^{r}(x,x',\omega)=\delta(x-x'),
\label{rgf1}
\end{equation}
After integrating Eq.~\eqref{rgf1} around $x=x'$ we find
\begin{equation}
\label{conditionGRSO}
[G^{r}(x>x')]_{x=x'}=[G^{r}(x<x')]_{x=x'}\,\,\, \mbox{and}\,\,\,
[\frac{d}{dx}G^{r}(x>x')]_{x=x'}-[\frac{d}{dx}G^{r}(x<x')]_{x=x'}=\eta\tau_{z}\sigma_{0},
\end{equation}
where $\tau_{i}$ represent Pauli matrices in particle-hole space, while $\sigma_{i}$ represent Pauli matrices in spin space and, $\eta=2m^{*}/\hbar^2$.
Generally, in particle-hole space $G^{r}$ is a $2\times2$ matrix,
\begin{equation}
\label{GF}
G^{r}(x,x',\omega)=
\begin{bmatrix}
G^{r}_{ee}&G^{r}_{eh}\\
G^{r}_{he}&G^{r}_{hh}
\end{bmatrix},
\end{equation}
where $G^{r}_{ee}$, $G^{r}_{eh}$, $G^{r}_{he}$ are matrix. In presence of spin flip scattering, we can write each element of $G^{r}(x,x',\omega)$ as
\begin{equation}
G_{ab}^{r}(x,x',\omega)=
\begin{pmatrix}
[G^{r}_{ab}]_{\uparrow\uparrow}&[G^{r}_{ab}]_{\uparrow\downarrow}\\
[G^{r}_{ab}]_{\downarrow\uparrow}&[G^{r}_{ab}]_{\downarrow\downarrow}
\end{pmatrix}, \mbox{with } a,b \in \{e,h\}.
\label{geh}
\end{equation}
Pairing amplitudes in our setup are calculated from retarded Green's function, as shown below.
\subsubsection{Pairing amplitudes}
The anomalous Green's function $G^{r}_{eh}$ is defined as,
\begin{equation}
G^{r}_{eh}(x,x',\omega)=i\sum_{\mu=0}^{3}f_{\mu}^{r}\sigma_{\mu}\sigma_{2},
\label{green}
\end{equation}
where $\sigma_{0}$ is a unit matrix, $\sigma_{\mu}(\mu=1,2,3)$ represent the Pauli matrices. $f_{0}^{r}$ represents SS ($\uparrow\downarrow-\downarrow\uparrow$), $f_{1,2}^{r}$ are the EST ($\downarrow\downarrow\pm\uparrow\uparrow$) and $f_{3}^{r}$ corresponds to the MST ($\uparrow\downarrow+\downarrow\uparrow$) components of the pairing amplitude in Eq.~\eqref{green}. The EST components $\uparrow\uparrow$ and $\downarrow\downarrow$ are given by $f_{\uparrow\uparrow}=if_{2}^{r}-f_{1}^{r}$ and $f_{\downarrow\downarrow}=if_{2}^{r}+f_{1}^{r}$, respectively. Even and odd-frequency components are calculated from,
\begin{equation}
\label{EVENODD}
f^{E}_{\mu}(x,x',\omega)=\frac{1}{2}[{f^{r}_{\mu}(x,x',\omega)+f^{a}_{\mu}(x,x',-\omega)}],\,\,\,\mbox{and}\,\,\,
f^{O}_{\mu}(x,x',\omega)=\frac{1}{2}[{f^{r}_{\mu}(x,x',\omega)-f^{a}_{\mu}(x,x',-\omega)}],
\end{equation}
$f_{\mu}^{a}$ being the advanced Green's function and can be derived using\cite{cayy} $G^{a}(x,x',\omega)=[G^{r}(x',x,\omega)]^{\dagger}$. The even and odd-frequency EST pairings can then be derived, see Eq.~\eqref{EVENODD}, as
\begin{equation}
\label{eo1}
\begin{split}
f_{\uparrow\uparrow}^{E}=if_{2}^{E}-f_{1}^{E},\hspace{1cm} f_{\downarrow\downarrow}^{E}=if_{2}^{E}+f_{1}^{E},\\
f_{\uparrow\uparrow}^{O}=if_{2}^{O}-f_{1}^{O},\hspace{1cm} f_{\downarrow\downarrow}^{O}=if_{2}^{O}+f_{1}^{O}.
\end{split}
\end{equation}
{In Appendix, we provide an explicit form of Green's functions.}
\section{Results}
{\subsection{YSR bound states in metal-spin flipper-metal-superconductor junction}
We calculate YSR bound states from the real part of the complex poles of differential charge conductance $\mathcal{G}_{c}$.}
In Figs.~2(a), (b) we plot YSR bound states as a function of interface transparency $Z$ for two cases: (a) when zero energy YSR states appear and (b) when zero energy YSR states don't appear. From Fig.~2(a), we see that for {low} values of spin flipper's spin and {high} values of exchange interaction, two bound state energies merge at zero energy for some particular values of $Z$. For these values of $Z$, zero energy YSR bound states appear as conductance peaks in Fig.~2(c) which depicts a plot of normalized charge conductance as an energy function. However, for other values of exchange interaction, no mergers are seen, see Fig.~2(b), implying the absence of zero energy YSR bound states.
\begin{figure}[h]
\centering{\includegraphics[width=.99\textwidth]{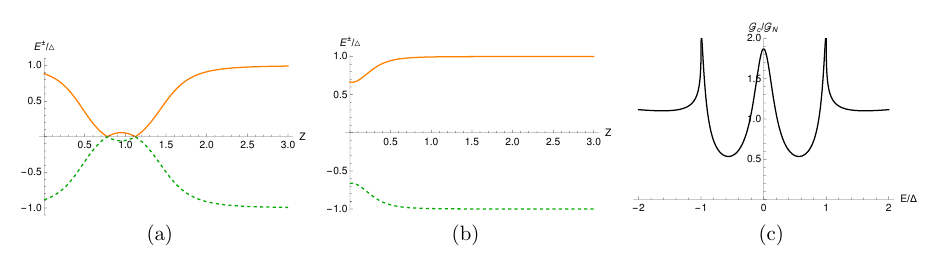}}
\caption{\small \sl Energy bound states vs $Z$ for (a) when zero energy YSR bound states appear and for (b) when zero energy YSR bound states don't appear, (c) Normalized charge conductance vs $E/\Delta$. Parameters: {$S=\frac{1}{2}$, $\mathcal{J}=4.5$ (for (a) and (c)), $\mathcal{J}=1.2$ (for (b)), $Z=0.776$ (for (c)), $k_{F}a=0.85\pi$ (for (a) and (c)), $k_{F}a=0.5\pi$ (for (b)). In (a) and (b) we consider the particular case of $S=-m'=1/2$ ($\mathcal{F}=1$). $\mathcal{F}=\sqrt{(S-m')(S+m'+1)}$ is the flip probability of spin flipper.}}
\end{figure}
{\subsection{YSR bound states in superconductor-metal-spin-flipper-metal-superconductor junction}
We compute Andreev energy bound states using the method as discussed in section II.A.7.} In Figs.~3(a), (b), we plot Andreev energy bound states as a function of exchange interaction $\mathcal{J}$ for {low values of spin flipper's spin}. From Fig.~3(a), we see that for {$\mathcal{J}=2.27$}, two bound state energies merge, and YSR bound states to occur; see also Ref.~\cite{costa}. However, from Fig.~3(b), it is seen that for {low} values of $\mathcal{J}$, no merger is seen, indicating the absence of zero energy YSR bound states. {In Fig.~3(c), we plot absolute value of Josephson current as a function of $\mathcal{J}$. We notice that for $\mathcal{J}=2.27$ when zero energy YSR bound states appear, there is a discontinuous change as a function of $\mathcal{J}$ in the absolute value of Josephson current. From the inset of Fig.~3(c), we see that Josephson current changes its sign (from positive to negative) at $\mathcal{J}=2.27$ because of a $0-\pi$ junction transition, marking the presence of a YSR state.}
\begin{figure}[h]
\centering{\includegraphics[width=.99\textwidth]{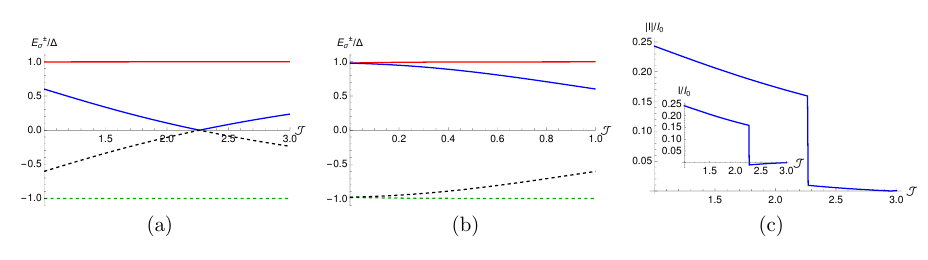}}
\caption{\small \sl (a) Energy bound states vs $\mathcal{J}$ in presence of spin flip scattering, (b) Energy bound states vs $\mathcal{J}$ in {presence} of spin flip scattering, (c) Josephson current and its absolute value vs $\mathcal{J}$. Parameters: {$S=\frac{1}{2}$, $\varphi=\pi/8$, $Z=0$, $k_{F}a=\pi/8$, $I_{0}=e\Delta/\hbar$. In (a) and (b) we consider the particular case of $S=-m'=1/2$ ($\mathcal{F}=1$).}}
\end{figure}
\subsection{Odd-frequency pairing and YSR bound states in N$_{1}$-SF-N$_{2}$-S junction}
\subsubsection{Spin-singlet superconducting pairing induced in N$_{1}$-SF-N$_{2}$-S junction}
We compute induced even-/odd-frequency superconducting pairing directly from anomalous components of $G^{r}(x,x',\omega)$ using Eqs.~\eqref{green}, \eqref{EVENODD} and \eqref{eo1}. For spin-singlet (SS) pairing we find,
{
\footnotesize
{
\begin{eqnarray}
\label{singleteven}
f_{0}^{E}(x,x',\omega)=&&-\frac{\eta\Delta e^{-\frac{\sqrt{\Delta^2-\omega^2}k_{F}}{2E_{F}}|x-x'|}}{4\sqrt{\Delta^2-\omega^2}}\Bigg[\frac{e^{ik_{F}|x-x'|}}{k_{F}+i\frac{\sqrt{\Delta^2-\omega^2}k_{F}}{2E_{F}}}+\frac{e^{-ik_{F}|x-x'|}}{k_{F}-i\frac{\sqrt{\Delta^2-\omega^2}k_{F}}{2E_{F}}}\Bigg]-\frac{\eta\Delta e^{-\frac{\sqrt{\Delta^2-\omega^2}k_{F}}{2E_{F}}|x-x'|}}{8\sqrt{\Delta^2-\omega^2}}\Bigg[\frac{(\it{b}_{51}+\it{b}_{62})e^{ik_{F}(x+x')}}{k_{F}+i\frac{\sqrt{\Delta^2-\omega^2}k_{F}}{2E_{F}}}+\nonumber\\
&&\frac{(\it{b}_{82}+\it{b}_{71})e^{-ik_{F}(x+x')}}{k_{F}-i\frac{\sqrt{\Delta^2-\omega^2}k_{F}}{2E_{F}}}\Bigg]-\frac{\eta e^{-\frac{\sqrt{\Delta^2-\omega^2}k_{F}}{2E_{F}}|x-x'|}}{4\sqrt{\Delta^2-\omega^2}}\frac{(\it{a}_{81}-\it{a}_{72})\cos[k_{F}(x-x')]\Big(\omega k_{F}-\big(\Delta^2-\omega^2\big)\frac{k_F}{2E_F}\Big)}{\Big(k_{F}^2+\big(\Delta^2-\omega^2\big)\frac{k_{F}^2}{4E_{F}^2}\Big)},\,\,\mbox{for}\,\, x>0\\
\label{singletodd}
\mbox{and}\,\,f_{0}^{O}(x,x',\omega)=&&\frac{\eta (\it{a}_{81}-\it{a}_{72})\Big(k_{F}+\frac{\omega k_{F}}{2E_F}\Big)}{4\Big(k_{F}^2+\big(\Delta^2-\omega^2\big)\frac{k_{F}^2}{4E_F^2}\Big)}\sin[k_{F}(x-x')]e^{-\frac{\sqrt{\Delta^2-\omega^2}k_{F}}{2E_{F}}(x+x')},\,\,\mbox{for}\,\, x>0,
\end{eqnarray}}}
\normalsize
where the normal ($\it{b}_{51}$, {$\it{b}_{62}$, $\it{b}_{71}$}, $\it{b}_{82}$) and Andreev reflection amplitudes ({$\it{a}_{72}$,} $\it{a}_{81}$) are calculated from the wavefunctions in Eq.~\eqref{wav}, and imposing boundary conditions on these, see Eqs.~\eqref{bc1}, \eqref{bc4}.
\begin{figure}[h]
\centering{\includegraphics[width=.99\textwidth]{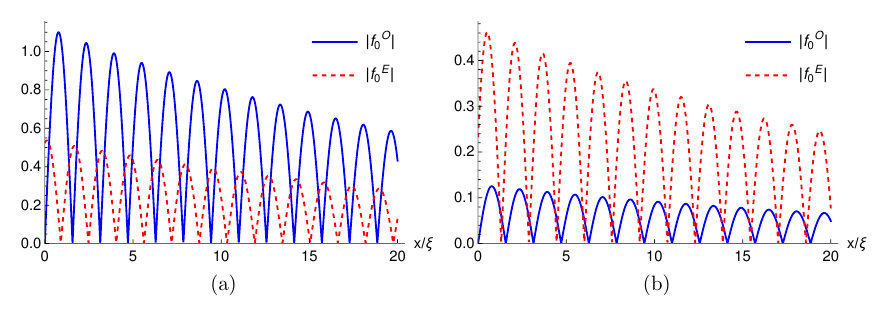}}
\caption{\small \sl The magnitudes of the Even-SS and Odd-SS pairings induced in the S region vs position $x$ for (a) when zero energy YSR states are present, (b) when zero energy YSR states are absent. Parameters: {$S=\frac{1}{2}$, $Z=0.776$, $\mathcal{J}=4.5$ (Fig.~4(a)), $\mathcal{J}=1.2$ (Fig.~4(b)), $x'=0$, $k_{F}\xi=2$, $k_{F}a=0.85\pi$ (Fig.~4(a)), $k_{F}a=0.5\pi$ (Fig.~4(b)), {$\omega\rightarrow0$}, $E_{F}=30\Delta$.}}
\end{figure}
{Bulk contribution to Even-SS pairing, from Eq.~\eqref{singleteven} is,
{
\begin{equation}
f_{0,B}^{E}=-\frac{\eta\Delta e^{-\frac{\sqrt{\Delta^2-\omega^2}k_{F}}{2E_{F}}|x-x'|}}{4\sqrt{\Delta^2-\omega^2}}\Bigg[\frac{e^{ik_{F}|x-x'|}}{k_{F}+i\frac{\sqrt{\Delta^2-\omega^2}k_{F}}{2E_{F}}}+\frac{e^{-ik_{F}|x-x'|}}{k_{F}-i\frac{\sqrt{\Delta^2-\omega^2}k_{F}}{2E_{F}}}\Bigg],
\end{equation}}
while {surface}  contributions from Eq.~\eqref{singleteven}, are
{
\begin{eqnarray}
f_{0,S}^{E}=&&-\frac{\eta\Delta e^{-\frac{\sqrt{\Delta^2-\omega^2}k_{F}}{2E_{F}}|x-x'|}}{8\sqrt{\Delta^2-\omega^2}}\Bigg[\frac{(\it{b}_{51}+\it{b}_{62})e^{ik_{F}(x+x')}}{k_{F}+i\frac{\sqrt{\Delta^2-\omega^2}k_{F}}{2E_{F}}}+\frac{(\it{b}_{82}+\it{b}_{71})e^{-ik_{F}(x+x')}}{k_{F}-i\frac{\sqrt{\Delta^2-\omega^2}k_{F}}{2E_{F}}}\Bigg]\nonumber\\
&&-\frac{\eta e^{-\frac{\sqrt{\Delta^2-\omega^2}k_{F}}{2E_{F}}|x-x'|}}{4\sqrt{\Delta^2-\omega^2}}\frac{(\it{a}_{81}-\it{a}_{72})\cos[k_{F}(x-x')]\Big(\omega k_{F}-\big(\Delta^2-\omega^2\big)\frac{k_F}{2E_F}\Big)}{\Big(k_{F}^2+\big(\Delta^2-\omega^2\big)\frac{k_{F}^2}{4E_{F}^2}\Big)}.
\end{eqnarray}}
Thus, Even-SS pairing has both bulk and {surface} components, while from Eq.~\eqref{singletodd}, we see that Odd-SS pairing has only {surface} component.

{To find the relation between surface odd-SS pairing and zero-energy YSR bound states, we focus on $\omega\rightarrow0$ limit. In this limit from Eqs.~\eqref{singleteven}, \eqref{singletodd}, we get,
\footnotesize
{
\begin{eqnarray}
\label{singletevenzero}
f_{0}^{E}(x,x',\omega\rightarrow0)=&&-\frac{\eta}{4}e^{-\frac{\Delta k_{F}}{2E_{F}}|x-x'|}\Bigg[\frac{e^{ik_{F}|x-x'|}}{k_{F}+i\frac{\Delta k_{F}}{2E_{F}}}+\frac{e^{-ik_{F}|x-x'|}}{k_{F}-i\frac{\Delta k_{F}}{2E_{F}}}\Bigg]-\frac{\eta}{8}e^{-\frac{\Delta k_{F}}{2E_{F}}|x-x'|}\Bigg[\frac{(\it{b}_{51}+\it{b}_{62})e^{ik_{F}(x+x')}}{k_{F}+i\frac{\Delta k_{F}}{2E_{F}}}+\frac{(\it{b}_{82}+\it{b}_{71})e^{-ik_{F}(x+x')}}{k_{F}-i\frac{\Delta k_{F}}{2E_{F}}}\Bigg]\nonumber\\
&&+\frac{\eta\Delta e^{-\frac{\Delta k_{F}}{2E_{F}}|x-x'|}}{8k_F E_F}\frac{(\it{a}_{81}-\it{a}_{72})\cos[k_{F}(x-x')]}{\Big(1+\frac{\Delta^2}{4E_{F}^2}\Big)},\,\,\mbox{for}\,\, x>0\\
\label{singletoddzero}
\mbox{and}\,\,f_{0}^{O}(x,x',\omega\rightarrow0)=&&\frac{\eta (\it{a}_{81}-\it{a}_{72})}{4\Big(k_{F}+\frac{\Delta^2k_{F}}{4E_F^2}\Big)}\sin[k_{F}(x-x')]e^{-\frac{\Delta k_{F}}{2E_{F}}(x+x')},\,\,\mbox{for}\,\, x>0.
\end{eqnarray}}}}
\normalsize From Eq.~\eqref{singletoddzero} we see that at $\omega\rightarrow0$, Odd-SS pairing depends on $\sin[k_{F}(x-x')]e^{-\gamma(x+x')}$, thus they exhibit an oscillatory decay with period $\frac{2\pi}{k_{F}}$ and decay length $\frac{1}{\gamma}$ where $\gamma=\sqrt{\Delta^2-\omega^2}[k_{F}/(2E_{F})]$.
Even-SS and Odd-SS superconducting pairings in the superconducting region {at $\omega\rightarrow0$} are plotted in Fig.~4. We consider two cases similar to Fig.~2. (a) when zero energy YSR states occur and (b) when zero energy YSR states are absent. From Fig.~4(a), we see that for {low} values of spin flipper spin and {high} values of exchange interaction $\mathcal{J}$, when zero energy YSR states appear (see Fig.~2(a)), {surface} Odd-SS pairing dominates Even-SS pairing. However, for {low} values of exchange interaction, for which zero energy YSR states are absent (see Fig.~2(b)), Even-SS pairing dominates {surface} Odd-SS pairing, as shown in Fig.~4(b). We conclude that the multi-fold enhancement of {surface} odd-frequency spin-singlet superconducting pairing is, therefore, due to the occurrence of zero energy YSR states.
\subsubsection{Spin-triplet superconducting pairing induced in N$_{1}$-SF-N$_{2}$-S junction}
ST pairing is two different types, EST and MST. However, we see MST pairing is absent and only EST pairing is finite since there is only spin flip scattering with vanishing spin mixing present in our setup (Fig.~1(a))\cite{odd}. For the Even-EST and Odd-EST superconducting pairing we get using Eq.~\eqref{eo1},
{
\begin{eqnarray}
\label{tripleteven}
f_{\uparrow\uparrow}^{E}(x,x',\omega)=&&-\frac{\eta (\it{a}_{62}-\it{a}_{51})\Big(k_{F}+\frac{\omega k_{F}}{2E_F}\Big)}{4\Big(k_{F}^2+\big(\Delta^2-\omega^2\big)\frac{k_{F}^2}{4E_F^2}\Big)}\sin[k_{F}(x-x')]e^{-\frac{\sqrt{\Delta^2-\omega^2}k_{F}}{2E_{F}}(x+x')}=-f_{\downarrow\downarrow}^{E}(x,x',\omega),\quad \mbox{for}\,\, x>0\\
\label{tripletodd}
f_{\uparrow\uparrow}^{O}(x,x',\omega)=&&-\frac{\eta\Delta e^{-\frac{\sqrt{\Delta^2-\omega^2}k_{F}}{2E_{F}}(x+x')}}{8\sqrt{\Delta^2-\omega^2}}\Bigg[\frac{(\it{b}_{72}+\it{b}_{81})e^{-ik_{F}(x+x')}}{k_{F}-i\frac{\sqrt{\Delta^2-\omega^2}k_{F}}{2E_{F}}}-\frac{(\it{b}_{61}+\it{b}_{52})e^{ik_{F}(x+x')}}{k_{F}+i\frac{\sqrt{\Delta^2-\omega^2}k_{F}}{2E_{F}}}\Bigg]\nonumber\\&&+\frac{\eta e^{-\frac{\sqrt{\Delta^2-\omega^2}k_{F}}{2E_{F}}(x+x')}}{4\sqrt{\Delta^2-\omega^2}}\frac{(\it{a}_{62}-\it{a}_{51})\cos[k_{F}(x-x')]\Big(\omega k_{F}-(\Delta^2-\omega^2)\frac{k_F}{2E_F}\Big)}{\Big(k_{F}^2+\big(\Delta^2-\omega^2\big)\frac{k_F^2}{4E_F^2}\Big)}\nonumber\\=&&-f_{\downarrow\downarrow}^{O}(x,x',\omega),\quad \mbox{for}\,\, x>0,
\end{eqnarray}}
where the normal ({$\it{b}_{52}$}, $\it{b}_{61}$, $\it{b}_{72}$, {$\it{b}_{81}$}) and Andreev reflection amplitudes ({$\it{a}_{51}$}, $\it{a}_{62}$) are calculated from the wavefunctions in Eq.~\eqref{wav} by imposing the boundary conditions Eqs.~\eqref{bc1}, \eqref{bc4}. {From Eqs.~\eqref{tripleteven}, \eqref{tripletodd}, we see that both Even-EST and Odd-EST pairings have only interface components.

{To understand the relation between surface odd-EST pairing and zero-energy YSR states, we concentrate on $\omega\rightarrow0$ limit. In this limit from Eqs.~\eqref{tripleteven}, \eqref{tripletodd}, we obtain,
\begin{eqnarray}
\label{tripletevenzero}
f_{\uparrow\uparrow}^{E}(x,x',\omega\rightarrow0)=&&-\frac{\eta (\it{a}_{62}-\it{a}_{51})}{4\Big(k_{F}+\frac{\Delta^2 k_{F}}{4E_F^2}\Big)}\sin[k_{F}(x-x')]e^{-\frac{\Delta k_{F}}{2E_{F}}(x+x')}=-f_{\downarrow\downarrow}^{E}(x,x',\omega\rightarrow0),\quad \mbox{for}\,\, x>0\\
\label{tripletoddzero}
f_{\uparrow\uparrow}^{O}(x,x',\omega\rightarrow0)=&&-\frac{\eta}{8}e^{-\frac{\Delta k_{F}}{2E_{F}}(x+x')}\Bigg[\frac{(\it{b}_{72}+\it{b}_{81})e^{-ik_{F}(x+x')}}{k_{F}-i\frac{\Delta k_{F}}{2E_{F}}}-\frac{(\it{b}_{61}+\it{b}_{52})e^{ik_{F}(x+x')}}{k_{F}+i\frac{\Delta k_{F}}{2E_{F}}}\Bigg]\nonumber\\&&-\frac{\eta\Delta e^{-\frac{\Delta k_{F}}{2E_{F}}(x+x')}}{8k_F E_F}\frac{(\it{a}_{62}-\it{a}_{51})\cos[k_{F}(x-x')]}{\Big(1+\frac{\Delta^2}{4E_F^2}\Big)}=-f_{\downarrow\downarrow}^{O}(x,x',\omega\rightarrow0),\quad \mbox{for}\,\, x>0.
\end{eqnarray}}
From Eq.~\eqref{tripletevenzero} we notice that at $\omega=0$, Even-EST superconducting pairing depends on $\sin[k_{F}(x-x')]e^{-\gamma(x+x')}$, showing an oscillatory decay with period $\frac{2\pi}{k_{F}}$. Even-EST and Odd-EST superconducting pairings as a function of $x$ in superconducting region {at $\omega\rightarrow0$} in presence of spin flip scattering are plotted in Fig.~5. We consider two cases: (a) when zero energy YSR states appear and, (b) when zero energy YSR states are absent. From Fig.~5(a) we notice that for parameters wherein zero energy YSR states occur, {surface} Odd-EST pairing is much larger than {surface} Even-EST pairing. {Surface} Even-EST pairing is almost vanishing.} However, when zero energy YSR states are absent, {surface} Even-EST pairing is much larger than {surface} Odd-EST pairing as shown in Fig.~5(b). Thus, our results indicate that {surface} Odd-EST pairing is enhanced considerably due to the occurrence of zero energy YSR states.
\begin{figure}[h]
\centering{\includegraphics[width=.99\textwidth]{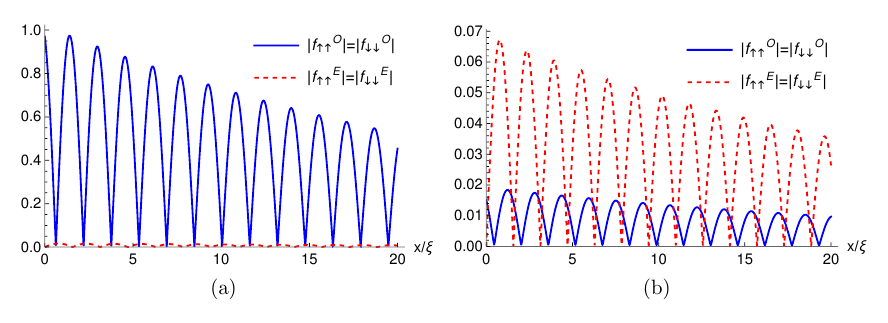}}
\caption{\small \sl The magnitudes of the Even-EST and Odd-EST pairing induced vs. position $x$ for (a) when zero energy YSR states occur and for (b) when zero energy YSR states are absent. Parameters: {$S=\frac{1}{2}$, $Z=0.776$, $\mathcal{J}=4.5$ (Fig.~5(a)), $\mathcal{J}=1.2$ (Fig.~5(b)), $x'=0$, $k_F\xi=2$, $k_{F}a=0.85\pi$ (Fig.~5(a)), $k_{F}a=0.5\pi$ (Fig.~5(b)), {$\omega\rightarrow0$}, $E_{F}=30\Delta$.}}
\end{figure}
{\subsubsection{Comparison between Even-SS and Odd-EST pairing in N$_{1}$-SF-N$_{2}$-S junction}
We compare Even-SS and Odd-EST pairing in metal-spin flipper-metal-Superconductor (N$_{1}$-SF-N$_{2}$-S) junction. In Fig.~6, we plot the Even-SS and Odd-EST superconducting pairings vs. $x$ in superconducting region {at $\omega\rightarrow0$} in case of N$_{1}$-SF-N$_{2}$-S junction. We consider two cases: (a) when zero energy YSR states are present and, (b) when zero energy YSR states are absent. From Fig.~6(a) we see that for parameters wherein zero energy YSR states are present, Odd-EST pairing is larger (around $\simeq2$ times) than Even-SS pairing. But, when zero energy YSR states are absent, Even-SS pairing is much larger (around $20$ times) than Odd-EST pairing as shown in Fig.~6(b).
\begin{figure}[h]
\centering{\includegraphics[width=.99\textwidth]{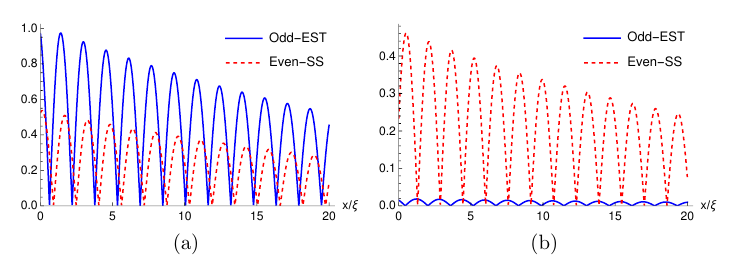}}
\caption{\small \sl The magnitudes of the induced Even-SS and Odd-EST pairing vs. position $x$ for (a) when zero energy YSR states occur and for (b) when zero energy YSR states are absent. Parameters: $S=\frac{1}{2}$, $Z=0.776$, $\mathcal{J}=4.5$ (Fig.~6(a)), $\mathcal{J}=1.2$ (Fig.~6(b)), $x'=0$, $E_{F}=30\Delta$, $k_F\xi=2$, $k_{F}a=0.85\pi$ (Fig.~6(a)), $k_{F}a=0.5\pi$ (Fig.~6(b)), {$\omega\rightarrow0$}.}
\end{figure}}

{In Fig.~7, we plot the Even-SS and Odd-EST superconducting pairings vs. frequency $\omega$. Similarly as before we consider two cases: (a) when zero energy YSR states occur and, (b) when zero energy YSR states are absent. We see that when zero energy YSR states appear, Odd-EST pairing develops a peak at $\omega=0$, and in the low-frequency regime, Odd-EST pairing is larger than Even-EST pairing. However, when zero energy YSR states are absent, Even-SS pairing is always larger than Odd-SS pairing.
\begin{figure}[h]
\centering{\includegraphics[width=.99\textwidth]{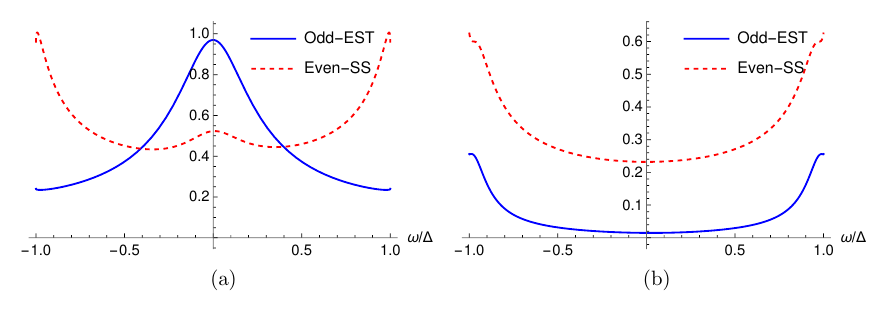}}
\caption{\small \sl The magnitudes of the induced Even-SS and Odd-EST pairing vs. frequency $\omega$ for (a) when zero energy YSR states occur and for (b) when zero energy YSR states are absent. Parameters: $S=\frac{1}{2}$, $Z=0.776$, $\mathcal{J}=4.5$ (Fig.~7(a)), $\mathcal{J}=1.2$ (Fig.~7(b)), $x=0$, $x'=0$, $E_{F}=30\Delta$, $k_F\xi=2$, $k_{F}a=0.85\pi$ (Fig.~7(a)), $k_{F}a=0.5\pi$ (Fig.~7(b)).}
\end{figure}}
\subsection{Odd-frequency pairing and YSR bound states in S-N$_{1}$-SF-N$_{2}$-S Josephson junction}
\subsubsection{Induced spin-singlet superconducting pairing in S-N$_{1}$-SF-N$_{2}$-S Josephson junction}
We compute even and odd-frequency SS pairing amplitudes from anomalous components of $G^{r}(x,x',\omega)$ using Eqs.~\eqref{green}, \eqref{EVENODD} and \eqref{eo1}. For Even-SS and Odd-SS superconducting pairing we get,
{
\footnotesize
{
\begin{eqnarray}
\label{ESE}
f_{0}^{E}(x,x',\omega)=&&-\frac{\eta\Delta e^{-\frac{\sqrt{\Delta^2-\omega^2}k_{F}}{2E_{F}}|x-x'|}}{4\sqrt{\Delta^2-\omega^2}}\Bigg[\frac{e^{ik_{F}|x-x'|}}{k_{F}+i\frac{\sqrt{\Delta^2-\omega^2}k_{F}}{2E_{F}}}+\frac{e^{-ik_{F}|x-x'|}}{k_{F}-i\frac{\sqrt{\Delta^2-\omega^2}k_{F}}{2E_{F}}}\Bigg]-\frac{\eta\Delta e^{\frac{\sqrt{\Delta^2-\omega^2}k_{F}}{2E_{F}}(x+x')}}{8\sqrt{\Delta^2-\omega^2}}\Bigg[\frac{(b'_{11}+b'_{22})e^{-ik_{F}(x+x')}}{k_{F}+i\frac{\sqrt{\Delta^2-\omega^2}k_{F}}{2E_{F}}}\nonumber\\
&&+\frac{(b'_{42}+b'_{31})e^{ik_{F}(x+x')}}{k_{F}-i\frac{\sqrt{\Delta^2-\omega^2}k_{F}}{2E_{F}}}\Bigg]-\frac{\eta e^{\frac{\sqrt{\Delta^2-\omega^2}k_{F}}{2E_{F}}(x+x')} }{8\sqrt{\Delta^2-\omega^2}}\cos[k_{F}(x-x')]\Bigg[\frac{(a'_{12}-a'_{21})\big(\omega-i\sqrt{\Delta^2-\omega^2}\big)}{k_{F}+i\frac{\sqrt{\Delta^2-\omega^2}k_{F}}{2E_{F}}}\nonumber\\&&+\frac{(a'_{41}-a'_{32})\big(\omega+i\sqrt{\Delta^2-\omega^2}\big)}{k_{F}-i\frac{\sqrt{\Delta^2-\omega^2}k_{F}}{2E_{F}}}\Bigg],\quad \mbox{for}\,\, x<0\\
f_{0}^{O}(x,x',\omega)=&&\frac{\eta\Big((a'_{12}-a'_{21})(\omega-i\sqrt{\Delta^2-\omega^2})\Big(k_{F}-i\frac{\sqrt{\Delta^2-\omega^2}k_{F}}{2E_{F}}\Big)-(a'_{41}-a'_{32})(\omega+i\sqrt{\Delta^2-\omega^2})\Big(k_{F}+i\frac{\sqrt{\Delta^2-\omega^2}k_{F}}{2E_{F}}\Big)\Big)}{8\sqrt{\Delta^2-\omega^2}\Big(k_{F}^2+(\Delta^2-\omega^2)\frac{k_F^2}{4E_F^2}\Big)}e^{\frac{\sqrt{\Delta^2-\omega^2}k_{F}}{2E_F}(x+x')}\nonumber\\&&\sin[k_{F}(x-x')],\quad \mbox{for}\,\, x<0.
\label{OSO}
\end{eqnarray}}}
\normalsize {Bulk contribution to Even-SS pairing, from Eq.~\eqref{ESE} is,
{
\begin{equation}
f_{0,B}^{E}=-\frac{\eta\Delta e^{-\frac{\sqrt{\Delta^2-\omega^2}k_{F}}{2E_{F}}|x-x'|}}{4\sqrt{\Delta^2-\omega^2}}\Bigg[\frac{e^{ik_{F}|x-x'|}}{k_{F}+i\frac{\sqrt{\Delta^2-\omega^2}k_{F}}{2E_{F}}}+\frac{e^{-ik_{F}|x-x'|}}{k_{F}-i\frac{\sqrt{\Delta^2-\omega^2}k_{F}}{2E_{F}}}\Bigg],
\end{equation}}
while {surface}  contributions from Eq.~\eqref{ESE}, are
{
\begin{eqnarray}
f_{0,S}^{E}=&&-\frac{\eta\Delta e^{\frac{\sqrt{\Delta^2-\omega^2}k_{F}}{2E_{F}}(x+x')}}{8\sqrt{\Delta^2-\omega^2}}\Bigg[\frac{(b'_{11}+b'_{22})e^{-ik_{F}(x+x')}}{k_{F}+i\frac{\sqrt{\Delta^2-\omega^2}k_{F}}{2E_{F}}}+\frac{(b'_{42}+b'_{31})e^{ik_{F}(x+x')}}{k_{F}-i\frac{\sqrt{\Delta^2-\omega^2}k_{F}}{2E_{F}}}\Bigg]-\frac{\eta e^{\frac{\sqrt{\Delta^2-\omega^2}k_{F}}{2E_{F}}(x+x')} }{8\sqrt{\Delta^2-\omega^2}}\cos[k_{F}(x-x')]\nonumber\\&&\Bigg[\frac{(a'_{12}-a'_{21})\big(\omega-i\sqrt{\Delta^2-\omega^2}\big)}{k_{F}+i\frac{\sqrt{\Delta^2-\omega^2}k_{F}}{2E_{F}}}+\frac{(a'_{41}-a'_{32})\big(\omega+i\sqrt{\Delta^2-\omega^2}\big)}{k_{F}-i\frac{\sqrt{\Delta^2-\omega^2}k_{F}}{2E_{F}}}\Bigg].
\end{eqnarray}}
Therefore, Even-SS pairing has both bulk and {surface} components, while from Eq.~\eqref{OSO}, we see that Odd-SS pairing has only {surface} component. Further,} from Eq.~\eqref{ESE} we notice that Even-SS superconducting pairing depends on both Andreev reflection ($a'_{12}$, {$a'_{21}$, $a'_{32}$,} $a'_{41}$ and normal reflection amplitudes ($b'_{11}$, {$b'_{22}$, $b'_{31}$, $b'_{42}$}), while in Eq.~\eqref{OSO} Odd-SS superconducting pairing depends only on Andreev reflection amplitudes ($a'_{12}$, {$a'_{21}$, $a'_{32}$,} $a'_{41}$). At $x=x'$ local Odd-SS pairing vanishes, while local Even-SS pairing is finite and is maximum.

{To examine the relation between surface odd-SS pairing and zero-energy YSR states, we focus on $\omega\rightarrow0$ limit. In this limit from Eqs.~\eqref{ESE}, \eqref{OSO}, we obtain,
\footnotesize
{
\begin{eqnarray}
\label{ESEzero}
f_{0}^{E}(x,x',\omega\rightarrow0)=&&-\frac{\eta}{4}e^{-\frac{\Delta k_{F}}{2E_{F}}|x-x'|}\Bigg[\frac{e^{ik_{F}|x-x'|}}{k_{F}+i\frac{\Delta k_{F}}{2E_{F}}}+\frac{e^{-ik_{F}|x-x'|}}{k_{F}-i\frac{\Delta k_{F}}{2E_{F}}}\Bigg]-\frac{\eta}{8}e^{\frac{\Delta k_{F}}{2E_{F}}(x+x')}\Bigg[\frac{(b'_{11}+b'_{22})e^{-ik_{F}(x+x')}}{k_{F}+i\frac{\Delta k_{F}}{2E_{F}}}+\frac{(b'_{42}+b'_{31})e^{ik_{F}(x+x')}}{k_{F}-i\frac{\Delta k_{F}}{2E_{F}}}\Bigg]\nonumber\\
&&+\frac{i\eta}{8}e^{\frac{\Delta k_{F}}{2E_{F}}(x+x')}\cos[k_{F}(x-x')]\Bigg[\frac{(a'_{12}-a'_{21})}{k_{F}+i\frac{\Delta k_{F}}{2E_{F}}}-\frac{(a'_{41}-a'_{32})}{k_{F}-i\frac{\Delta k_{F}}{2E_{F}}}\Bigg],\quad \mbox{for}\,\, x<0\\
f_{0}^{O}(x,x',\omega\rightarrow0)=&&-\frac{i\eta\Big((a'_{12}-a'_{21})\Big(1-i\frac{\Delta}{2E_{F}}\Big)+(a'_{41}-a'_{32})\Big(1+i\frac{\Delta}{2E_{F}}\Big)\Big)}{8\Big(k_{F}+\frac{\Delta^2 k_F}{4E_F^2}\Big)}e^{\frac{\Delta k_{F}}{2E_F}(x+x')}\sin[k_{F}(x-x')],\quad \mbox{for}\,\, x<0.
\label{OSOzero}
\end{eqnarray}}}
In Fig.~8, we plot Even-SS and Odd-SS superconducting pairings induced in left superconducting and metallic regions {at $\omega\rightarrow0$}. We consider two cases: (a) when zero energy YSR states occur and (b) when zero energy YSR states are absent. From Fig.~8(a), we see that for {$\mathcal{J}=2.27$} when zero energy YSR states occur (see Fig.~3(a)), {surface} Odd-SS pairing is larger than Even-SS pairing. However, for {$\mathcal{J}=0.5$}, when zero energy YSR states are absent (see Fig.~3(b)), Even-SS pairing is much more prominent while {surface} Odd-SS pairing vanishes as shown in Fig.~8(b). Thus, in the presence of YSR states, there is an enhancement of {surface} Odd-SS pairing magnitude.
\begin{figure}[h]
\centering{\includegraphics[width=.99\textwidth]{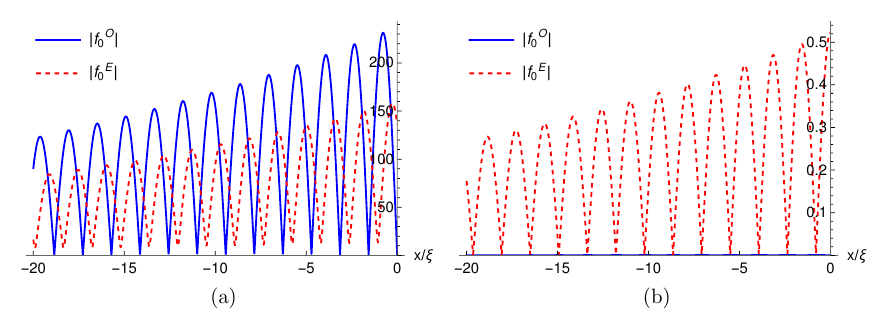}}
\caption{\small \sl The magnitudes of the induced Even-SS and Odd-SS pairing vs. position $x$ for (a) when zero energy YSR states occur and for (b) when zero energy YSR states are absent. Parameters: {$S=\frac{1}{2}$, $Z=0$, $\mathcal{J}=2.27$ (Fig.~8(a)), $\mathcal{J}=0.5$ (Fig.~8(b)), $x'=0$, $E_{F}=30\Delta$, $k_F\xi=2$, $k_{F}a=\pi/8$, {$\omega\rightarrow0$,} $\varphi=\pi/8$, $T=0$.}}
\end{figure}
\subsubsection{Induced spin-triplet superconducting pairing in S-N$_{1}$-SF-N$_{2}$-S Josephson junction}
Next, we compute the induced Even-ST and Odd-ST pairing in S-N$_{1}$-SF-N$_{2}$-S junction. In this paper, as mentioned before MST pairing vanishes and only EST pairing is finite as there is absence of spin mixing in our setup, see {section IV and} Ref.~\cite{odd} for the reasons behind it. Using Eq.~\eqref{eo1}, the induced Even-EST and Odd-EST superconducting pairings are,
{
\small
{
\begin{eqnarray}
\label{ETO}
f_{\uparrow\uparrow}^{E}(x,x',\omega)=&&\frac{\eta e^{\frac{\sqrt{\Delta^2-\omega^2}k_{F}}{2E_{F}}(x+x')}}{8\sqrt{\Delta^2-\omega^2}}\sin[k_{F}(x-x')]\Bigg[\frac{(a'_{11}-a'_{22})\big(\omega-\sqrt{\Delta^2-\omega^2}\big)}{k_{F}+i\frac{\sqrt{\Delta^2-\omega^2}k_{F}}{2E_{F}}}+\frac{(a'_{42}-a'_{31})\big(\omega+\sqrt{\Delta^2-\omega^2}\big)}{k_{F}-i\frac{\sqrt{\Delta^2-\omega^2}k_{F}}{2E_{F}}}\Bigg]\nonumber\\=&&-f_{\downarrow\downarrow}^{E}(x,x',\omega),\,\,\mbox{for}\,x<0\\
f_{\uparrow\uparrow}^{O}(x,x',\omega)=&&\frac{\eta\Delta e^{\frac{\sqrt{\Delta^2-\omega^2}k_{F}}{2E_{F}}(x+x')}}{8\sqrt{\Delta^2-\omega^2}}\Bigg[\frac{(b'_{12}+b'_{21})e^{-ik_{F}(x+x')}}{k_{F}+i\frac{\sqrt{\Delta^2-\omega^2}k_{F}}{2E_{F}}}-\frac{(b'_{41}+b'_{32})e^{ik_{F}(x+x')}}{k_{F}-i\frac{\sqrt{\Delta^2-\omega^2}k_{F}}{2E_{F}}}\Bigg]\nonumber\\
&&-\frac{\eta e^{\frac{\sqrt{\Delta^2-\omega^2}k_{F}}{2E_{F}}(x+x')}}{4\sqrt{\Delta^2-\omega^2}}\cos[k_{F}(x-x')]\Bigg[\frac{(a'_{11}-a'_{22})\big(\omega-\sqrt{\Delta^2-\omega^2}\big)}{k_{F}+i\frac{\sqrt{\Delta^2-\omega^2}k_{F}}{2E_{F}}}-\frac{(a'_{42}-a'_{31})\big(\omega+\sqrt{\Delta^2-\omega^2}\big)}{k_{F}-i\frac{\sqrt{\Delta^2-\omega^2}k_{F}}{2E_{F}}}\Bigg]\nonumber\\=&&-f_{\downarrow\downarrow}^{O}(x,x',\omega),\quad \mbox{for}\,\, x<0.
\label{OTE}
\end{eqnarray}}}
\normalsize {From Eqs.~\eqref{ETO},\eqref{OTE}, we see that both Even-EST and Odd-EST pairings have only {surface} components. Further,} from Eqs.~\eqref{ETO} and, \eqref{OTE} we notice that the induced Even-EST pairing depends only on Andreev reflection amplitudes ($a'_{11}$, {$a'_{22}$, $a'_{31}$}, $a'_{42}$), while the induced Odd-EST pairing depends on both normal ($b'_{12}$, {$b'_{21}$, $b'_{32}$,} $b'_{41}$) as well as Andreev reflection amplitudes ($a'_{11}$, {$a'_{22}$, $a'_{31}$}, $a'_{42}$). At $x=x'$, the induced Even-EST pairing vanishes, while the induced Odd-EST pairing is finite.

{To explore the connection between surface odd-EST pairing and zero-energy YSR states, we focus on the limit as $\omega\rightarrow0$. In this limit from Eqs.~\eqref{ETO}, \eqref{OTE}, we get,
\small
{
\begin{eqnarray}
\label{ETOzero}
f_{\uparrow\uparrow}^{E}(x,x',\omega\rightarrow0)=&&\frac{\eta}{8}e^{\frac{\Delta k_{F}}{2E_{F}}(x+x')}\sin[k_{F}(x-x')]\Bigg[-\frac{(a'_{11}-a'_{22})}{k_{F}+i\frac{\Delta k_{F}}{2E_{F}}}+\frac{(a'_{42}-a'_{31})}{k_{F}-i\frac{\Delta k_{F}}{2E_{F}}}\Bigg]=-f_{\downarrow\downarrow}^{E}(x,x',\omega\rightarrow0),\,\,\mbox{for}\,x<0\\
f_{\uparrow\uparrow}^{O}(x,x',\omega\rightarrow0)=&&\frac{\eta}{8}e^{\frac{\Delta k_{F}}{2E_{F}}(x+x')}\Bigg[\frac{(b'_{12}+b'_{21})e^{-ik_{F}(x+x')}}{k_{F}+i\frac{\Delta k_{F}}{2E_{F}}}-\frac{(b'_{41}+b'_{32})e^{ik_{F}(x+x')}}{k_{F}-i\frac{\Delta k_{F}}{2E_{F}}}\Bigg]\nonumber\\
&&+\frac{\eta}{4}e^{\frac{\Delta k_{F}}{2E_{F}}(x+x')}\cos[k_{F}(x-x')]\Bigg[\frac{(a'_{11}-a'_{22})}{k_{F}+i\frac{\Delta k_{F}}{2E_{F}}}+\frac{(a'_{42}-a'_{31})}{k_{F}-i\frac{\Delta k_{F}}{2E_{F}}}\Bigg]=-f_{\downarrow\downarrow}^{O}(x,x',\omega\rightarrow0),\quad \mbox{for}\,\, x<0.
\label{OTEzero}
\end{eqnarray}}}
In Fig.~9, we plot the induced Even-EST and Odd-EST superconducting pairings in left superconducting and metallic regions {at $\omega\rightarrow0$} vs. position $x$. Similar to Fig.~4, we consider two cases: (a) when zero energy YSR states occur and (b) when zero energy YSR states are absent. From Fig.~9(a), we notice that for {$\mathcal{J}=2.27$} when zero energy YSR states occur, the induced {surface} Odd-EST pairing dominates over {surface} Even-EST pairing. However, for {$\mathcal{J}=0.5$} when zero energy YSR states are absent, the induced {surface} Even-EST pairing is much larger than the {surface} Odd-EST pairing, as seen from Fig.~9(b). This suggests a boost to {surface} Odd-EST pairing due to zero energy YSR bound states. {The magnitude of {surface} Odd-EST pairing is large near the spin flipper location at $x=0$ when YSR bound states occur, however, far away from $x=0$, the magnitude of {surface} Odd-EST pairing is small. The nature of {surface} Odd-EST pairing is maximum at spin flipper's location, it always exhibits an oscillatory decay.}
\begin{figure}[h]
\centering{\includegraphics[width=.99\textwidth]{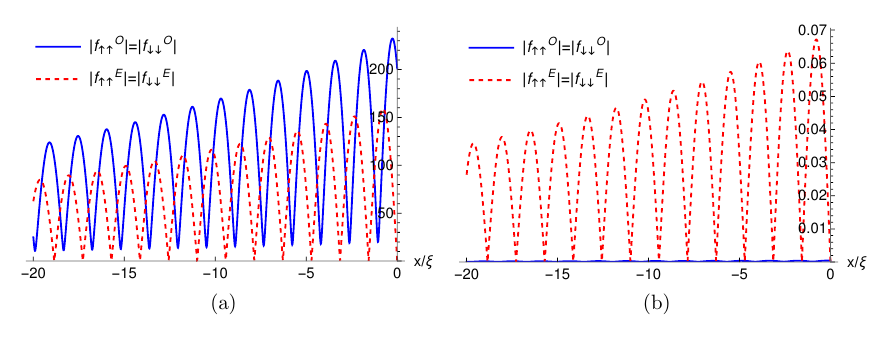}}
\caption{\small \sl The magnitudes of the induced Even-EST and Odd-EST superconductor pairing in the left superconducting and normal metal regions vs position $x$ for (a) when zero energy YSR states occur and for (b) when zero energy YSR states are absent. Parameters: $S=\frac{1}{2}$, $Z=0$, $\mathcal{J}=2.27$ (Fig.~9(a)), $\mathcal{J}=0.5$ (Fig.~9(b)), $x'=0$, $E_{F}=30\Delta$, $k_{F}a=\pi/8$, $k_{F}\xi=2$, {$\omega\rightarrow0$}, $\varphi=\pi/8$, $T=0$.}
\end{figure}
{\subsubsection{Comparison between Even-SS and Odd-EST pairing in S-N$_{1}$-SF-N$_{2}$-S junction}
We compare Even-SS and Odd-EST pairing in Superconductor-metal-spin flipper-metal-Superconductor (S-N$_{1}$-SF-N$_{2}$-S) junction. In Fig.~10, we plot the induced Even-SS and Odd-EST superconducting pairing in left superconducting and normal metal regions vs. $x$ in case of S-N$_{1}$-SF-N$_{2}$-S junction. Similarly as before, we consider two cases: (a) when zero energy YSR states are present and, (b) when zero energy YSR states are absent. From Fig.~10(a), we see that for $\mathcal{J}=2.27$ when zero energy YSR states arise, the induced Odd-EST pairing is larger than Even-SS pairing. However, for $J=0.5$ when zero energy YSR states are absent, the induced Even-SS pairing only exists with vanishing Odd-EST pairing.
\begin{figure}[h]
\centering{\includegraphics[width=0.99\textwidth]{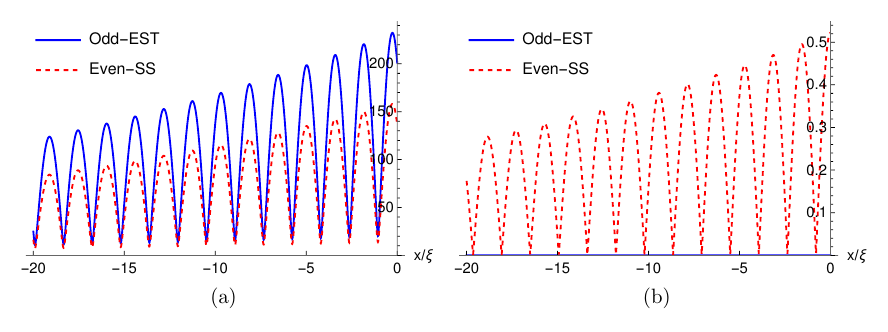}}
\caption{\small \sl The magnitudes of the induced Even-SS and Odd-EST superconductor pairing in the left superconducting and normal metal regions vs position $x$ for (a) when zero energy YSR states occur and for (b) when zero energy YSR states are absent. Parameters: $S=\frac{1}{2}$, $Z=0$, $\mathcal{J}=2.27$ (Fig.~10(a)), $\mathcal{J}=0.5$ (Fig.~10(b)), $x'=0$, $E_{F}=30\Delta$, $k_{F}a=\pi/8$, $k_F\xi=2$, {$\omega\rightarrow0$,} $\varphi=\pi/8$, $T=0$.}
\end{figure}}
\begin{figure}[h]
\centering{\includegraphics[width=.99\textwidth]{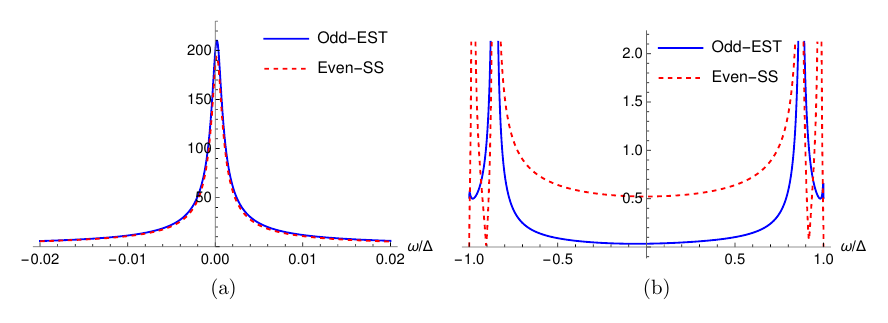}}
\caption{\small \sl The magnitudes of the induced Even-SS and Odd-EST pairing vs. frequency $\omega$ for (a) when zero energy YSR states occur and for (b) when zero energy YSR states are absent. Parameters: $S=\frac{1}{2}$, $Z=0$, $\mathcal{J}=2.27$ (Fig.~11(a)), $\mathcal{J}=0.5$ (Fig.~11(b)), $x=0$, $x'=0$, $E_{F}=30\Delta$, $k_F\xi=2$, $k_{F}a=\pi/8$, $\varphi=\pi/8$.}
\end{figure}

{In Fig.~11 the Even-SS and Odd-EST superconducting pairings are plotted as function of frequency $\omega$. We consider two cases: (a) when zero energy YSR states appear and, (b) when zero energy YSR states are absent. We notice that in presence of YSR states Odd-EST pairing is larger than Even-SS pairing. But, when YSR states are absent, Even-SS pairing is much larger than Odd-SS pairing.}
{The effect of interface transparency and finite temperature on superconducting pairing is discussed in Appendix.}

\section{Analysis}
Table I compares our results when zero energy YSR states occur and when they are absent in $N_{1}-sf-N_{2}-S$ and $S-N_{1}-sf-N_{2}-S$ junctions.  When zero energy YSR states are present, an almost quantized zero-bias peak in the conductance spectra is observed, while no such peak is present in their absence. The induced odd-frequency superconducting pairing is enhanced multi-fold over even frequency pairing due to the presence of YSR states. This enhancement of odd-frequency pairing is explained by the interaction of the spin flipper's spin with the Andreev reflected electrons or holes, which gives rise to YSR bound states at the NS interface. The YSR states break the translational symmetry and mix spatial parities, inducing Odd-SS pairing at the interface. In addition, spin-flip scattering induces Odd-EST pairing. Thus, both Andreev reflection and spin-flip scattering are responsible for odd-frequency pairing as well as YSR bound states in the setup. {When YSR states emerge at zero energy, it creates an opportunity for odd-frequency Cooper pairs to form. This is because, in the presence of YSR states,
there is a non-zero amplitude for odd-frequency Cooper pairs to exist at zero energy, resulting in an
enhancement of odd-frequency pairing correlations.}
For $S-N_{1}-sf-N_{2}-S$ Josephson junction, the presence of YSR bound states leads to a $0-\pi$ junction transition and an enhanced odd-frequency pairing. In contrast, in the absence of YSR states, no $0-\pi$ junction transition occurs, and even frequency superconducting pairing is much more prominent while odd-frequency pairing vanishes. Overall, the presence of YSR bound states in the two setups has significant effects on the conductance spectra and induced superconducting pairing, with an enhancement of odd-frequency pairing being a signature of YSR bound states. {Therefore,  in a system exhibiting both YSR states and odd-frequency pairing, via observing odd-frequency pairing, one can detect the presence of YSR bound states. Experimentally, the evidence of odd-frequency pairing in heterostructures can be observed by measuring long-range supercurrents in ferromagnetic Josephson junctions\cite{ber,khai}.}
\begin{table}[h]
\caption{Comparing induced odd $\omega$ pairing in presence and absence of zero energy YSR states}
\begin{tabular}{|p{1.5cm}|p{6.5cm}|p{6.5cm}|}
\hline
& Zero energy YSR states occur & Zero energy YSR states are absent\\
\hline
\multirow{3}{*}{N-sf-N-S} & (1) Almost quantized zero bias conductance peak. & (1) No zero bias conductance peak.
\\\cline{2-3}{}{}
& (2) Induced odd-frequency pairing (both singlet and equal spin-triplet) dominates even frequency pairing. & (2) Induced even frequency pairing (both singlet and equal spin-triplet) dominates odd-frequency pairing.\\
\hline\hline
\multirow{3}{*}{S-N-sf-N-S} & (1) There is a $0-\pi$ junction transition when zero energy YSR states occur. & (1) There is no $0-\pi$ junction transition.
\\\cline{2-3}{}{}
& (2) Odd-frequency superconducting pairing {(both singlet and equal spin-triplet)} is {larger than} even frequency superconducting pairing. & (2) Even frequency superconducting pairing {(both singlet and equal spin-triplet)} is much larger with vanishing odd-frequency superconducting pairing.\\
\hline
\end{tabular}
\end{table}

{To understand the reasons why spin-flip scattering engenders EST pairing with vanishing MST pairing, we look at other spin based scattering mechanisms possible at such junctions. A particular spin based scattering mechanism which can occur is spin mixing. Spin mixing and spin-flip scattering very different processes. In the spin mixing process, an electron encounters spin-dependent phase shifts\cite{lin}, while in the spin-flip scattering process, spin of the electron flips\cite{AJP}. For  example, when an electron transmits through a Ferromagnet-Superconductor (FS) junction only spin mixing happens\cite{dutt}, while when an electron transmits through a Ferromagnet-Ferromagnet-Superconductor (F$_{1}$F$_{2}$S) junction where the magnetization vectors are misaligned, both spin mixing and spin flip scattering happen\cite{meng}. However, when an electron scatters from the spin flipper as in our work only spin-flip scattering occurs. Thus we have three spin based scattering mechanisms which may occur : (1) only spin mixing, (2) only spin flip scattering  and, (3) spin mixing and spin flip scattering both coexist. For spin mixing only happens, odd- and even-frequency MST pairings exist but EST pairing is zero. For spin flip scattering only happens, odd- and even-frequency EST pairings are finite but MST pairing is zero. However, when spin mixing and spin flip scattering both happen,  even- as well as odd-frequency EST and MST pairings are finite, see section V of Ref.~\cite{odd} for more details.}

\section{Experimental realization and Conclusion}
As shown in Fig.~1, our setups can be experimentally realized easily since the metal-superconductor junctions were realized in a lab around 45 years ago\cite{gus}. Placing a spin flipper at the metal-superconductor interface should not be cumbersome; particularly with a conventional superconductor, it should be possible. {Experimentally, high spin molecules like Fe$_{19}$-complex\cite{ara} with a spin of $S=33/2$ or Mn$_{4}$O$_{3}$ complex\cite{wer} with a spin of $S=9/2$ can to a certain extent be a model for the spin flipper. The internal dynamics of such a high spin molecule are perhaps quite distinct from the spin-flipper. However, the spin flipper can mimic the half-integer spin states up to any arbitrary high value and the associated spin magnetic moment of the high spin molecule and the consequence of an electron interacting with such, to a large extent. A thin ferromagnetic layer can not be used as a spin-flipper since it does not generate any spin-flip scattering; it only generate spin mixing\cite{moz}. In our work, delta potential acts as a system parameter only, it has nothing to do with spin degree of freedom. So if the spin-flip layer is thicker or if one use a thin rectangular potential barrier instead of delta potential, main conclusion of our work does not change. Exchange coupling of spin flipper can be tuned experimentally through mechanical forces applied by an scanning tunneling microscopy (STM) tip\cite{fari}.}

We conclude with the main findings of this work. This work identifies a strong correlation between {surface odd-frequency pairing} and the presence of zero energy YSR states in both metal-spin flipper-metal-$s$-wave superconductor junction and $s$-wave superconductor-metal-spin flipper-metal-$s$-wave superconductor junction. When zero energy YSR states are present, the induced {surface} odd-frequency pairing is much larger than even-frequency pairing. Conversely, when zero energy YSR states are absent, even-frequency pairing dominates {surface} odd-frequency pairing. In the case of $s$-wave superconductor-metal-spin flipper-metal-$s$-wave superconductor junction, the occurrence of zero energy YSR states results in a $0-\pi$ junction transition, and {surface} odd-frequency pairing dominates even-frequency pairing. These findings suggest that large {surface} odd-frequency pairing serves as a fingerprint for the presence of YSR states.
\acknowledgments This work was supported by the grants: 1. Josephson junctions with strained Dirac materials and their application in quantum information processing, SERB Grant No. CRG/20l9/006258, and 2. Nash equilibrium versus Pareto optimality in N-Player games, SERB MATRICS Grant No. MTR/2018/000070.

\appendix
{\section{Wavefunctions in metal-spin flipper-metal-superconductor junction}
The wavefunctions in different domains of metal-spin flipper-metal-superconductor junction for various kinds of scattering processes are given as-
\footnotesize{
\begin{eqnarray}
\label{wav}
\begin{split}
{}\varphi_{1}(x)=\begin{cases}
\chi_{1}^{N}e^{iq_{e}(x+a)}\phi_{m'}^{S}+\it{a}_{11}\chi_{3}^{N}e^{iq_{h}(x+a)}\phi_{m'+1}^{S}+\it{a}_{12}\chi_{4}^{N}e^{iq_{h}(x+a)}\phi_{m'}^{S}+\it{b}_{11}\chi_{1}^{N}e^{-iq_{e}(x+a)}\phi_{m'}^{S}+\it{b}_{12}\chi_{2}^{N}e^{-iq_{e}(x+a)}\phi_{m'+1}^{S}, \hspace{0.5cm}\text{$x<-a$}\\
\it{c}_{11}\chi_{1}^{N}e^{iq_{e}(x+a)}\phi_{m'}^{S}+\it{c}_{12}\chi_{2}^{N}e^{iq_{e}(x+a)}\phi_{m'+1}^{S}+\it{d}_{11}\chi_{1}^{N}e^{-iq_{e}x}\phi_{m'}^{S}+\it{d}_{12}\chi_{2}^{N}e^{-iq_{e}x}\phi_{m'+1}^{S}+\it{e}_{11}\chi_{3}^{N}e^{iq_{h}x}\phi_{m'+1}^{S}+\it{e}_{12}\chi_{4}^{N}e^{iq_{h}x}\phi_{m'}^{S}\\
+\it{f}_{11}\chi_{3}^{N}e^{-iq_{h}(x+a)}\phi_{m'+1}^{S}+\it{f}_{12}\chi_{4}^{N}e^{-iq_{h}(x+a)}\phi_{m'}^{S},\hspace{9cm}\text{$-a<x<0$}\\
\it{g}_{11}\chi_{1}^{S}e^{iq_{e}^{S}x}\phi_{m'}^{S}+\it{g}_{12}\chi_{2}^{S}e^{iq_{e}^{S}x}\phi_{m'+1}^{S}+\it{h}_{11}\chi_{3}^{S}e^{-iq_{h}^{S}x}\phi_{m'+1}^{S}+\it{h}_{12}\chi_{4}^{S}e^{-iq_{h}^{S}x}\phi_{m'}^{S}, \hspace{5.2cm}\text{$x>0$}
\end{cases}\\
{}\varphi_{2}(x)=\begin{cases}
\chi_{2}^{N}e^{iq_{e}(x+a)}\phi_{m'}^{S}+\it{a}_{21}\chi_{3}^{N}e^{iq_{h}(x+a)}\phi_{m'}^{S}+\it{a}_{22}\chi_{4}^{N}e^{iq_{h}(x+a)}\phi_{m'-1}^{S}+b_{21}\chi_{1}^{N}e^{-iq_{e}(x+a)}\phi_{m'-1}^{S}+\it{b}_{22}\chi_{2}^{N}e^{-iq_{e}(x+a)}\phi_{m'}^{S}, \hspace{0.2cm}\text{$x<-a$}\\
\it{c}_{21}\chi_{1}^{N}e^{iq_{e}(x+a)}\phi_{m'-1}^{S}+\it{c}_{22}\chi_{2}^{N}e^{iq_{e}(x+a)}\phi_{m'}^{S}+\it{d}_{21}\chi_{1}^{N}e^{-iq_{e}x}\phi_{m'-1}^{S}+\it{d}_{22}\chi_{2}^{N}e^{-iq_{e}x}\phi_{m'}^{S}+\it{e}_{21}\chi_{3}^{N}e^{iq_{h}x}\phi_{m'}^{S}+\it{e}_{22}\chi_{4}^{N}e^{iq_{h}x}\phi_{m'-1}^{S}\\
+\it{f}_{21}\chi_{3}^{N}e^{-iq_{h}(x+a)}\phi_{m'}^{S}+\it{f}_{22}\chi_{4}^{N}e^{-iq_{h}(x+a)}\phi_{m'-1}^{S},\hspace{9cm}\text{$-a<x<0$}\\
\it{g}_{21}\chi_{1}^{S}e^{iq_{e}^{S}x}\phi_{m'-1}^{S}+\it{g}_{22}\chi_{2}^{S}e^{iq_{e}^{S}x}\phi_{m'}^{S}+\it{h}_{21}\chi_{3}^{S}e^{-iq_{h}^{S}x}\phi_{m'}^{S}+\it{h}_{22}\chi_{4}^{S}e^{-iq_{h}^{S}x}\phi_{m'-1}^{S}, \hspace{5.2cm}\text{$x>0$}
\end{cases}\\
{}\varphi_{3}(x)=\begin{cases}
\chi_{3}^{N}e^{-iq_{h}(x+a)}\phi_{m'}^{S}+\it{a}_{31}\chi_{1}^{N}e^{iq_{e}(x+a)}\phi_{m'-1}^{S}+\it{a}_{32}\chi_{2}^{N}e^{iq_{e}(x+a)}\phi_{m'}^{S}+\it{b}_{31}\chi_{3}^{N}e^{-iq_{h}(x+a)}\phi_{m'}^{S}+\it{b}_{32}\chi_{4}^{N}e^{-iq_{h}(x+a)}\phi_{m'-1}^{S}, \hspace{0.2cm}\text{$x<-a$}\\
\it{c}_{31}\chi_{1}^{N}e^{iq_{e}(x+a)}\phi_{m'-1}^{S}+\it{c}_{32}\chi_{2}^{N}e^{iq_{e}(x+a)}\phi_{m'}^{S}+\it{d}_{31}\chi_{1}^{N}e^{-iq_{e}x}\phi_{m'-1}^{S}+\it{d}_{32}\chi_{2}^{N}e^{-iq_{e}x}\phi_{m'}^{S}+\it{e}_{31}\chi_{3}^{N}e^{iq_{h}x}\phi_{m'}^{S}+\it{e}_{32}\chi_{4}^{N}e^{iq_{h}x}\phi_{m'-1}^{S}\\
+\it{f}_{31}\chi_{3}^{N}e^{-iq_{h}(x+a)}\phi_{m'}^{S}+\it{f}_{32}\chi_{4}^{N}e^{-iq_{h}(x+a)}\phi_{m'-1}^{S},\hspace{9cm}\text{$-a<x<0$}\\
\it{g}_{31}\chi_{1}^{S}e^{iq_{e}^{S}x}\phi_{m'-1}^{S}+\it{g}_{32}\chi_{2}^{S}e^{iq_{e}^{S}x}\phi_{m'}^{S}+\it{h}_{31}\chi_{3}^{S}e^{-iq_{h}^{S}x}\phi_{m'}^{S}+\it{h}_{32}\chi_{4}^{S}e^{-iq_{h}^{S}x}\phi_{m'-1}^{S}, \hspace{5.2cm}\text{$x>0$}
\end{cases}\\
{}\varphi_{4}(x)=\begin{cases}
\chi_{4}^{N}e^{-iq_{h}(x+a)}\phi_{m'}^{S}+\it{a}_{41}\chi_{1}^{N}e^{iq_{e}(x+a)}\phi_{m'}^{S}+\it{a}_{42}\chi_{2}^{N}e^{iq_{e}(x+a)}\phi_{m'+1}^{S}+\it{b}_{41}\chi_{3}^{N}e^{-iq_{h}(x+a)}\phi_{m'+1}^{S}+\it{b}_{42}\chi_{4}^{N}e^{-iq_{h}(x+a)}\phi_{m'}^{S}, \hspace{0.2cm}\text{$x<-a$}\\
\it{c}_{41}\chi_{1}^{N}e^{iq_{e}(x+a)}\phi_{m'}^{S}+\it{c}_{42}\chi_{2}^{N}e^{iq_{e}(x+a)}\phi_{m'+1}^{S}+\it{d}_{41}\chi_{1}^{N}e^{-iq_{e}x}\phi_{m'}^{S}+\it{d}_{42}\chi_{2}^{N}e^{-iq_{e}x}\phi_{m'+1}^{S}+\it{e}_{41}\chi_{3}^{N}e^{iq_{h}x}\phi_{m'+1}^{S}+\it{e}_{42}\chi_{4}^{N}e^{iq_{h}x}\phi_{m'}^{S}\\
+\it{f}_{41}\chi_{3}^{N}e^{-iq_{h}(x+a)}\phi_{m'+1}^{S}+\it{f}_{42}\chi_{4}^{N}e^{-iq_{h}(x+a)}\phi_{m'}^{S},\hspace{9cm}\text{$-a<x<0$}\\
\it{g}_{41}\chi_{1}^{S}e^{iq_{e}^{S}x}\phi_{m'}^{S}+\it{g}_{42}\chi_{2}^{S}e^{iq_{e}^{S}x}\phi_{m'+1}^{S}+\it{h}_{41}\chi_{3}^{S}e^{-iq_{h}^{S}x}\phi_{m'+1}^{S}+\it{h}_{42}\chi_{4}^{S}e^{-iq_{h}^{S}x}\phi_{m'}^{S}, \hspace{5.2cm}\text{$x>0$}
\end{cases}\\
{}\varphi_{5}(x)=\begin{cases}
\it{g}_{51}\chi_{1}^{N}e^{iq_{e}(x+a)}\phi_{m'}^{S}+\it{g}_{52}\chi_{2}^{N}e^{iq_{e}(x+a)}\phi_{m'+1}^{S}+\it{h}_{51}\chi_{3}^{N}e^{-iq_{h}(x+a)}\phi_{m'+1}^{S}+\it{h}_{52}\chi_{4}^{N}e^{-iq_{h}(x+a)}\phi_{m'}^{S}, \hspace{3.35cm}\text{$x<-a$}\\
\it{c}_{51}\chi_{1}^{N}e^{iq_{e}(x+a)}\phi_{m'}^{S}+\it{c}_{52}\chi_{2}^{N}e^{iq_{e}(x+a)}\phi_{m'+1}^{S}+\it{d}_{51}\chi_{1}^{N}e^{-iq_{e}x}\phi_{m'}^{S}+\it{d}_{52}\chi_{2}^{N}e^{-iq_{e}x}\phi_{m'+1}^{S}+\it{e}_{51}\chi_{3}^{N}e^{iq_{h}x}\phi_{m'+1}^{S}+\it{e}_{52}\chi_{4}^{N}e^{iq_{h}x}\phi_{m'}^{S}\\
+\it{f}_{51}\chi_{3}^{N}e^{-iq_{h}(x+a)}\phi_{m'+1}^{S}+\it{f}_{52}\chi_{4}^{N}e^{-iq_{h}(x+a)}\phi_{m'}^{S},\hspace{9cm}\text{$-a<x<0$}\\
\chi_{1}^{S}e^{-iq_{e}^{S}x}\phi_{m'}^{S}+\it{a}_{51}\chi_{3}^{S}e^{-iq_{h}^{S}x}\phi_{m'+1}^{S}+\it{a}_{52}\chi_{4}^{S}e^{-iq_{h}^{S}x}\phi_{m'}^{S}+\it{b}_{51}\chi_{1}^{S}e^{iq_{e}^{S}x}\phi_{m'}^{S}+\it{b}_{52}\chi_{2}^{S}e^{iq_{e}^{S}x}\phi_{m'+1}^{S}, \hspace{3cm}\text{$x>0$}
\end{cases}\\
{}\varphi_{6}(x)=\begin{cases}
\it{g}_{61}\chi_{1}^{N}e^{iq_{e}(x+a)}\phi_{m'-1}^{S}+\it{g}_{62}\chi_{2}^{N}e^{iq_{e}(x+a)}\phi_{m'}^{S}+\it{h}_{61}\chi_{3}^{N}e^{-iq_{h}(x+a)}\phi_{m'}^{S}+\it{h}_{62}\chi_{4}^{N}e^{-iq_{h}(x+a)}\phi_{m'-1}^{S}, \hspace{3.35cm}\text{$x<-a$}\\
\it{c}_{61}\chi_{1}^{N}e^{iq_{e}(x+a)}\phi_{m'-1}^{S}+\it{c}_{62}\chi_{2}^{N}e^{iq_{e}(x+a)}\phi_{m'}^{S}+\it{d}_{61}\chi_{1}^{N}e^{-iq_{e}x}\phi_{m'-1}^{S}+\it{d}_{62}\chi_{2}^{N}e^{-iq_{e}x}\phi_{m'}^{S}+\it{e}_{61}\chi_{3}^{N}e^{iq_{h}x}\phi_{m'}^{S}+\it{e}_{62}\chi_{4}^{N}e^{iq_{h}x}\phi_{m'-1}^{S}\\
+\it{f}_{61}\chi_{3}^{N}e^{-iq_{h}(x+a)}\phi_{m'}^{S}+\it{f}_{62}\chi_{4}^{N}e^{-iq_{h}(x+a)}\phi_{m'-1}^{S},\hspace{9cm}\text{$-a<x<0$}\\
\chi_{2}^{S}e^{-iq_{e}^{S}x}\phi_{m'}^{S}+\it{a}_{61}\chi_{3}^{S}e^{-iq_{h}^{S}x}\phi_{m'}^{S}+\it{a}_{62}\chi_{4}^{S}e^{-iq_{h}^{S}x}\phi_{m'-1}^{S}+\it{b}_{61}\chi_{1}^{S}e^{iq_{e}^{S}x}\phi_{m'-1}^{S}+\it{b}_{62}\chi_{2}^{S}e^{iq_{e}^{S}x}\phi_{m'}^{S},\hspace{3cm}\text{$x>0$}
\end{cases}\\
{}\varphi_{7}(x)=\begin{cases}
\it{g}_{71}\chi_{1}^{N}e^{iq_{e}(x+a)}\phi_{m'-1}^{S}+\it{g}_{72}\chi_{2}^{N}e^{iq_{e}(x+a)}\phi_{m'}^{S}+\it{h}_{71}\chi_{3}^{N}e^{-iq_{h}(x+a)}\phi_{m'}^{S}+\it{h}_{72}\chi_{4}^{N}e^{-iq_{h}(x+a)}\phi_{m'-1}^{S},\hspace{3.35cm} \text{$x<-a$}\\
\it{c}_{71}\chi_{1}^{N}e^{iq_{e}(x+a)}\phi_{m'-1}^{S}+\it{c}_{72}\chi_{2}^{N}e^{iq_{e}(x+a)}\phi_{m'}^{S}+\it{d}_{71}\chi_{1}^{N}e^{-iq_{e}x}\phi_{m'-1}^{S}+\it{d}_{72}\chi_{2}^{N}e^{-iq_{e}x}\phi_{m'}^{S}+\it{e}_{71}\chi_{3}^{N}e^{iq_{h}x}\phi_{m'}^{S}+\it{e}_{72}\chi_{4}^{N}e^{iq_{h}x}\phi_{m'-1}^{S}\\
+\it{f}_{71}\chi_{3}^{N}e^{-iq_{h}(x+a)}\phi_{m'}^{S}+\it{f}_{72}\chi_{4}^{N}e^{-iq_{h}(x+a)}\phi_{m'-1}^{S},\hspace{9cm}\text{$-a<x<0$}\\
\chi_{3}^{S}e^{iq_{h}^{S}x}\phi_{m'}^{S}+\it{a}_{71}\chi_{1}^{S}e^{-iq_{e}^{S}x}\phi_{m'-1}^{S}+\it{a}_{72}\chi_{2}^{S}e^{-iq_{e}^{S}x}\phi_{m'}^{S}+\it{b}_{71}\chi_{3}^{S}e^{iq_{h}^{S}x}\phi_{m'}^{S}+\it{b}_{72}\chi_{4}^{S}e^{iq_{h}^{S}x}\phi_{m'-1}^{S}, \hspace{3cm}\text{$x>0$}
\end{cases}\\
{}\varphi_{8}(x)=\begin{cases}
\it{g}_{81}\chi_{1}^{N}e^{iq_{e}(x+a)}\phi_{m'}^{S}+\it{g}_{82}\chi_{2}^{N}e^{iq_{e}(x+a)}\phi_{m'+1}^{S}+\it{h}_{81}\chi_{3}^{N}e^{-iq_{h}(x+a)}\phi_{m'+1}^{S}+\it{h}_{82}\chi_{4}^{N}e^{-iq_{h}(x+a)}\phi_{m'}^{S},\hspace{3.35cm} \text{$x<-a$}\\
\it{c}_{81}\chi_{1}^{N}e^{iq_{e}(x+a)}\phi_{m'}^{S}+\it{c}_{82}\chi_{2}^{N}e^{iq_{e}(x+a)}\phi_{m'+1}^{S}+\it{d}_{81}\chi_{1}^{N}e^{-iq_{e}x}\phi_{m'}^{S}+\it{d}_{82}\chi_{2}^{N}e^{-iq_{e}x}\phi_{m'+1}^{S}+\it{e}_{81}\chi_{3}^{N}e^{iq_{h}x}\phi_{m'+1}^{S}+\it{e}_{82}\chi_{4}^{N}e^{iq_{h}x}\phi_{m'}^{S}\\
+\it{f}_{81}\chi_{3}^{N}e^{-iq_{h}(x+a)}\phi_{m'+1}^{S}+\it{f}_{82}\chi_{4}^{N}e^{-iq_{h}(x+a)}\phi_{m'}^{S},\hspace{9cm}\text{$-a<x<0$}\\
\chi_{4}^{S}e^{iq_{h}^{S}x}\phi_{m'}^{S}+\it{a}_{81}\chi_{1}^{S}e^{-iq_{e}^{S}x}\phi_{m'}^{S}+\it{a}_{82}\chi_{2}^{S}e^{-iq_{e}^{S}x}\phi_{m'+1}^{S}+\it{b}_{81}\chi_{3}^{S}e^{iq_{h}^{S}x}\phi_{m'+1}^{S}+\it{b}_{82}\chi_{4}^{S}e^{iq_{h}^{S}x}\phi_{m'}^{S}, \hspace{3cm}\text{$x>0$}
\end{cases}
\end{split}
\end{eqnarray}}
\normalsize
where $\chi_{1}^{N}=\begin{bmatrix}
1\\
0\\
0\\
0
\end{bmatrix}$, $\chi_{2}^{N}=\begin{bmatrix}
0\\
1\\
0\\
0
\end{bmatrix}$, $\chi_{3}^{N}=\begin{bmatrix}
0\\
0\\
1\\
0
\end{bmatrix}$, $\chi_{4}^{N}=\begin{bmatrix}
0\\
0\\
0\\
1
\end{bmatrix}$, $\chi_{1}^{S}=\begin{bmatrix}
u\\
0\\
0\\
v
\end{bmatrix}$, $\chi_{2}^{S}=\begin{bmatrix}
0\\
-u\\
v\\
0
\end{bmatrix}$, $\chi_{3}^{S}=\begin{bmatrix}
0\\
-v\\
u\\
0
\end{bmatrix}$ and $\chi_{4}^{S}=\begin{bmatrix}
v\\
0\\
0\\
u
\end{bmatrix}$. In Eq.~\eqref{wav}, $\varphi_{1}$, $\varphi_{2}$, $\varphi_{3}$, and $\varphi_{4}$ represent the wavefunctions when spin up electron, spin down electron, spin up hole and spin down hole are injected from left metallic region respectively, while $\varphi_{5}$, $\varphi_{6}$, $\varphi_{7}$, and $\varphi_{8}$ represent the wavefunctions when spin up electronlike quasiparticle, spin down electronlike quasiparticle, spin up holelike quasiparticle and spin down holelike quasiparticle are injected from superconductor respectively. $\it{a}_{ij}$ and $\it{b}_{ij}$ represent Andreev and normal reflection coefficient respectively, while $\it{g}_{ij}$ and $\it{h}_{ij}$ represent transmission coefficients for electron/electronlike and hole/holelike quasiparticles respectively. The spin flipper's eigenfunction is represented by $\phi_{m'}^{S}$ with its spin $S$ and spin magnetic moment $m'$. The $z$ component of spin flipper's spin $S^{z}$ is acting as- $S^{z}\phi_{m'}^{S}=\hbar m'\phi_{m'}^{S}$. $u=\sqrt{\frac{1}{2}\big(\frac{\omega+\sqrt{\omega^2-\Delta^2}}{\omega}\big)}$ and $v=\sqrt{\frac{1}{2}\big(\frac{\omega-\sqrt{\omega^2-\Delta^2}}{\omega}\big)}$ are the BCS coherence factors. $q_{e,h}=\sqrt{\frac{2m^{*}}{\hbar^2}(E_{F}\pm\omega)}$ is the wave-vector in normal metal, while $q_{e,h}^{S}=\sqrt{\frac{2m^{*}}{\hbar^2}(E_{F}\pm\sqrt{\omega^2-\Delta^2})}$ is the wave-vector in superconductor. If we diagonalize the Hamiltonian $(H_{BdG}^{\mbox{N$_{1}$-sf-N$_{2}$-S}})^{*}(-k)$ instead of $H_{BdG}^{\mbox{N$_{1}$-sf-N$_{2}$-S}}(k)$ we will get conjugated processes $\tilde{\varphi_{i}}$ which is necessary to form the retarded Green's functions in next subsection. For our setup (Fig.~1) we note that $\tilde{\chi_{i}}^{N(S)}=\chi_{i}^{N(S)}$. In our work we consider $E_{F}\gg\
\Delta,\omega$, in this limit $q_{e,h}\approx k_{F}(1\pm\frac{\omega}{2E_{F}})$ and $q_{e,h}^{S}\approx k_{F}\pm i\gamma$ where $k_{F}=\sqrt{\frac{2m^{*}E_{F}}{\hbar^2}}$ and $\gamma=\sqrt{\Delta^2-\omega^2}[k_{F}/(2E_{F})]$. $\xi=\frac{\hbar v_{F}}{\Delta}$ is the superconducting coherence length\cite{smp}.
\section{Wavefunctions in superconductor-metal-spin flipper-metal-superconductor junction}
The wavefunctions in different domains of superconductor-metal-spin flipper-metal-superconductor junction for various kinds of scattering processes are given as-
\footnotesize{
\begin{eqnarray}
\begin{split}
{}\varphi_{1}(x)=\begin{cases}
\chi_{1}^{S}e^{iq_{e}^{S}x}\phi_{m'}^{S}+a'_{11}\chi_{3}^{S}e^{iq_{h}^{S}x}\phi_{m'+1}^{S}+a'_{12}\chi_{4}^{S}e^{iq_{h}^{S}x}\phi_{m'}^{S}+b'_{11}\chi_{1}^{S}e^{-iq_{e}^{S}x}\phi_{m'}^{S}+b'_{12}\chi_{2}^{S}e^{-iq_{e}^{S}x}\phi_{m'+1}^{S}, \hspace{0.5cm}\text{$x<-a/2$},\\
c'_{11}\chi_{1}^{N}e^{iq_{e}(x+a/2)}\phi_{m'}^{S}+c'_{12}\chi_{2}^{N}e^{iq_{e}(x+a/2)}\phi_{m'+1}^{S}+d'_{11}\chi_{1}^{N}e^{-iq_{e}x}\phi_{m'}^{S}+d'_{12}\chi_{2}^{N}e^{-iq_{e}x}\phi_{m'+1}^{S}+e'_{11}\chi_{3}^{N}e^{iq_{h}x}\phi_{m'+1}^{S}+e'_{12}\chi_{4}^{N}e^{iq_{h}x}\phi_{m'}^{S}\\
+f'_{11}\chi_{3}^{N}e^{-iq_{h}(x+a/2)}\phi_{m'+1}^{S}+f'_{12}\chi_{4}^{N}e^{-iq_{h}(x+a/2)}\phi_{m'}^{S},\hspace{6.8cm}\text{$-a/2<x<0$},\\
g'_{11}\chi_{1}^{N}e^{-iq_{e}(x-a/2)}\phi_{m'}^{S}+g'_{12}\chi_{2}^{N}e^{-iq_{e}(x-a/2)}\phi_{m'+1}^{S}+h'_{11}\chi_{1}^{N}e^{iq_{e}x}\phi_{m'}^{S}+h'_{12}\chi_{2}^{N}e^{iq_{e}x}\phi_{m'+1}^{S}+i'_{11}\chi_{3}^{N}e^{-iq_{h}x}\phi_{m'+1}^{S}+i'_{12}\chi_{4}^{N}e^{-iq_{h}x}\phi_{m'}^{S}\\
+j'_{11}\chi_{3}^{N}e^{iq_{h}(x-a/2)}\phi_{m'+1}^{S}+j'_{12}\chi_{4}^{N}e^{iq_{h}(x-a/2)}\phi_{m'}^{S},\hspace{7.5cm}\text{$0<x<a/2$},\\
k'_{11}\chi_{1}^{S_{R}}e^{iq_{e}^{S}x}\phi_{m'}^{S}+k'_{12}\chi_{2}^{S_{R}}e^{iq_{e}^{S}x}\phi_{m'+1}^{S}+l'_{11}\chi_{3}^{S_{R}}e^{-iq_{h}^{S}x}\phi_{m'+1}^{S}+l'_{12}\chi_{4}^{S_{R}}e^{-iq_{h}^{S}x}\phi_{m'}^{S}, \hspace{3.1cm}\text{$x>a/2$}.
\end{cases}\\
{}\varphi_{2}(x)=\begin{cases}
\chi_{2}^{S}e^{iq_{e}^{S}x}\phi_{m'}^{S}+a'_{21}\chi_{3}^{S}e^{iq_{h}^{S}x}\phi_{m'}^{S}+a'_{22}\chi_{4}^{S}e^{iq_{h}^{S}x}\phi_{m'-1}^{S}+b'_{21}\chi_{1}^{S}e^{-iq_{e}^{S}x}\phi_{m'-1}^{S}+b'_{22}\chi_{2}^{S}e^{-iq_{e}^{S}x}\phi_{m'}^{S}, \hspace{0.5cm}\text{$x<-a/2$},\\
c'_{21}\chi_{1}^{N}e^{iq_{e}(x+a/2)}\phi_{m'-1}^{S}+c'_{22}\chi_{2}^{N}e^{iq_{e}(x+a/2)}\phi_{m'}^{S}+d'_{21}\chi_{1}^{N}e^{-iq_{e}x}\phi_{m'-1}^{S}+d'_{22}\chi_{2}^{N}e^{-iq_{e}x}\phi_{m'}^{S}+e'_{21}\chi_{3}^{N}e^{iq_{h}x}\phi_{m'}^{S}+e'_{22}\chi_{4}^{N}e^{iq_{h}x}\phi_{m'-1}^{S}\\
+f'_{21}\chi_{3}^{N}e^{-iq_{h}(x+a/2)}\phi_{m'}^{S}+f'_{22}\chi_{4}^{N}e^{-iq_{h}(x+a/2)}\phi_{m'-1}^{S},\hspace{6.8cm}\text{$-a/2<x<0$},\\
g'_{21}\chi_{1}^{N}e^{-iq_{e}(x-a/2)}\phi_{m'-1}^{S}+g'_{22}\chi_{2}^{N}e^{-iq_{e}(x-a/2)}\phi_{m'}^{S}+h'_{21}\chi_{1}^{N}e^{iq_{e}x}\phi_{m'-1}^{S}+h'_{22}\chi_{2}^{N}e^{iq_{e}x}\phi_{m'}^{S}+i'_{21}\chi_{3}^{N}e^{-iq_{h}x}\phi_{m'}^{S}+i'_{22}\chi_{4}^{N}e^{-iq_{h}x}\phi_{m'-1}^{S}\\
+j'_{21}\chi_{3}^{N}e^{iq_{h}(x-a/2)}\phi_{m'}^{S}+j'_{22}\chi_{4}^{N}e^{iq_{h}(x-a/2)}\phi_{m'-1}^{S},\hspace{7.5cm}\text{$0<x<a/2$},\\
k'_{21}\chi_{1}^{S_{R}}e^{iq_{e}^{S}x}\phi_{m'-1}^{S}+k'_{22}\chi_{2}^{S_{R}}e^{ik_{e}^{S}x}\phi_{m'}^{S}+l'_{21}\chi_{3}^{S_{R}}e^{-ik_{h}^{S}x}\phi_{m'}^{S}+l'_{22}\chi_{4}^{S_{R}}e^{-ik_{h}^{S}x}\phi_{m'-1}^{S}, \hspace{3.1cm}\text{$x>a/2$}.
\end{cases}\\
{}\varphi_{3}(x)=\begin{cases}
\chi_{3}^{S}e^{-iq_{h}^{S}x}\phi_{m'}^{S}+b'_{31}\chi_{3}^{S}e^{iq_{h}^{S}x}\phi_{m'}^{S}+b'_{32}\chi_{4}^{S}e^{iq_{h}^{S}x}\phi_{m'-1}^{S}+a'_{31}\chi_{1}^{S}e^{-iq_{e}^{S}x}\phi_{m'-1}^{S}+a'_{32}\chi_{2}^{S}e^{-iq_{e}^{S}x}\phi_{m'}^{S}, \hspace{0.5cm}\text{$x<-a/2$},\\
c'_{31}\chi_{1}^{N}e^{iq_{e}(x+a/2)}\phi_{m'-1}^{S}+c'_{32}\chi_{2}^{N}e^{iq_{e}(x+a/2)}\phi_{m'}^{S}+d'_{31}\chi_{1}^{N}e^{-iq_{e}x}\phi_{m'-1}^{S}+d'_{32}\chi_{2}^{N}e^{-iq_{e}x}\phi_{m'}^{S}+e'_{31}\chi_{3}^{N}e^{iq_{h}x}\phi_{m'}^{S}+e'_{32}\chi_{4}^{N}e^{iq_{h}x}\phi_{m'-1}^{S}\\
+f'_{31}\chi_{3}^{N}e^{-iq_{h}(x+a/2)}\phi_{m'}^{S}+f'_{32}\chi_{4}^{N}e^{-iq_{h}(x+a/2)}\phi_{m'-1}^{S},\hspace{6.8cm}\text{$-a/2<x<0$},\\
g'_{31}\chi_{1}^{N}e^{-iq_{e}(x-a/2)}\phi_{m'-1}^{S}+g'_{32}\chi_{2}^{N}e^{-iq_{e}(x-a/2)}\phi_{m'}^{S}+h'_{31}\chi_{1}^{N}e^{iq_{e}x}\phi_{m'-1}^{S}+h'_{32}\chi_{2}^{N}e^{iq_{e}x}\phi_{m'}^{S}+i'_{31}\chi_{3}^{N}e^{-iq_{h}x}\phi_{m'}^{S}+i'_{32}\chi_{4}^{N}e^{-iq_{h}x}\phi_{m'-1}^{S}\\
+j'_{31}\chi_{3}^{N}e^{iq_{h}(x-a/2)}\phi_{m'}^{S}+j'_{32}\chi_{4}^{N}e^{iq_{h}(x-a/2)}\phi_{m'-1}^{S},\hspace{7.5cm}\text{$0<x<a/2$},\\
k'_{31}\chi_{1}^{S_{R}}e^{iq_{e}^{S}x}\phi_{m'-1}^{S}+k'_{32}\chi_{2}^{S_{R}}e^{iq_{e}^{S}x}\phi_{m'}^{S}+l'_{31}\chi_{3}^{S_{R}}e^{-iq_{h}^{S}x}\phi_{m'}^{S}+l'_{32}\chi_{4}^{S_{R}}e^{-iq_{h}^{S}x}\phi_{m'-1}^{S}, \hspace{3.1cm}\text{$x>a/2$}.
\end{cases}\\
{}\varphi_{4}(x)=\begin{cases}
\chi_{4}^{S}e^{-iq_{h}^{S}x}\phi_{m'}^{S}+b'_{41}\chi_{3}^{S}e^{iq_{h}^{S}x}\phi_{m'+1}^{S}+b'_{42}\chi_{4}^{S}e^{iq_{h}^{S}x}\phi_{m'}^{S}+a'_{41}\chi_{1}^{S}e^{-iq_{e}^{S}x}\phi_{m'}^{S}+a'_{42}\chi_{2}^{S}e^{-iq_{e}^{S}x}\phi_{m'+1}^{S}, \hspace{0.5cm}\text{$x<-a/2$},\\
c'_{41}\chi_{1}^{N}e^{iq_{e}(x+a/2)}\phi_{m'}^{S}+c'_{42}\chi_{2}^{N}e^{iq_{e}(x+a/2)}\phi_{m'+1}^{S}+d'_{41}\chi_{1}^{N}e^{-iq_{e}x}\phi_{m'}^{S}+d'_{42}\chi_{2}^{N}e^{-iq_{e}x}\phi_{m'+1}^{S}+e'_{41}\chi_{3}^{N}e^{iq_{h}x}\phi_{m'+1}^{S}+e'_{42}\chi_{4}^{N}e^{iq_{h}x}\phi_{m'}^{S}\\
+f'_{41}\chi_{3}^{N}e^{-iq_{h}(x+a/2)}\phi_{m'+1}^{S}+f'_{42}\chi_{4}^{N}e^{-iq_{h}(x+a/2)}\phi_{m'}^{S},\hspace{6.8cm}\text{$-a/2<x<0$},\\
g'_{41}\chi_{1}^{N}e^{-iq_{e}(x-a/2)}\phi_{m'}^{S}+g'_{42}\chi_{2}^{N}e^{-iq_{e}(x-a/2)}\phi_{m'+1}^{S}+h'_{41}\chi_{1}^{N}e^{iq_{e}x}\phi_{m'}^{S}+h'_{42}\chi_{2}^{N}e^{iq_{e}x}\phi_{m'+1}^{S}+i'_{41}\chi_{3}^{N}e^{-iq_{h}x}\phi_{m'+1}^{S}+i'_{42}\chi_{4}^{N}e^{-iq_{h}x}\phi_{m'}^{S}\\
+j'_{41}\chi_{3}^{N}e^{iq_{h}(x-a/2)}\phi_{m'+1}^{S}+j'_{42}\chi_{4}^{N}e^{iq_{h}(x-a/2)}\phi_{m'}^{S},\hspace{7.5cm}\text{$0<x<a/2$},\\
k'_{41}\chi_{1}^{S_{R}}e^{iq_{e}^{S}x}\phi_{m'}^{S}+k'_{42}\chi_{2}^{S_{R}}e^{iq_{e}^{S}x}\phi_{m'+1}^{S}+l'_{41}\chi_{3}^{S_{R}}e^{-iq_{h}^{S}x}\phi_{m'+1}^{S}+l'_{42}\chi_{4}^{S_{R}}e^{-iq_{h}^{S}x}\phi_{m'}^{S}, \hspace{3.1cm}\text{$x>a/2$}.
\end{cases}\\
{}\varphi_{5}(x)=\begin{cases}
k'_{51}\chi_{1}^{S}e^{-iq_{e}^{S}x}\phi_{m'}^{S}+k'_{52}\chi_{2}^{S}e^{-iq_{e}^{S}x}\phi_{m'+1}^{S}+l'_{51}\chi_{3}^{S}e^{iq_{h}^{S}x}\phi_{m'+1}^{S}+l'_{52}\chi_{4}^{S}e^{iq_{h}^{S}x}\phi_{m'}^{S}, \hspace{3.5cm}\text{$x<-a/2$},\\
c'_{51}\chi_{1}^{N}e^{iq_{e}(x+a/2)}\phi_{m'}^{S}+c'_{52}\chi_{2}^{N}e^{iq_{e}(x+a/2)}\phi_{m'+1}^{S}+d'_{51}\chi_{1}^{N}e^{-iq_{e}x}\phi_{m'}^{S}+d'_{52}\chi_{2}^{N}e^{-iq_{e}x}\phi_{m'+1}^{S}+e'_{51}\chi_{3}^{N}e^{iq_{h}x}\phi_{m'+1}^{S}+e'_{52}\chi_{4}^{N}e^{iq_{h}x}\phi_{m'}^{S}\\
+f'_{51}\chi_{3}^{N}e^{-iq_{h}(x+a/2)}\phi_{m'+1}^{S}+f'_{52}\chi_{4}^{N}e^{-iq_{h}(x+a/2)}\phi_{m'}^{S},\hspace{6.8cm}\text{$-a/2<x<0$},\\
g'_{51}\chi_{1}^{N}e^{-iq_{e}(x-a/2)}\phi_{m'}^{S}+g'_{52}\chi_{2}^{N}e^{-iq_{e}(x-a/2)}\phi_{m'+1}^{S}+h'_{51}\chi_{1}^{N}e^{iq_{e}x}\phi_{m'}^{S}+h'_{52}\chi_{2}^{N}e^{iq_{e}x}\phi_{m'+1}^{S}+i'_{51}\chi_{3}^{N}e^{-iq_{h}x}\phi_{m'+1}^{S}+i'_{52}\chi_{4}^{N}e^{-iq_{h}x}\phi_{m'}^{S}\\
+j'_{51}\chi_{3}^{N}e^{iq_{h}(x-a/2)}\phi_{m'+1}^{S}+j'_{52}\chi_{4}^{N}e^{iq_{h}(x-a/2)}\phi_{m'}^{S},\hspace{7.5cm}\text{$0<x<a/2$},\\
\chi_{1}^{S_{R}}e^{-iq_{e}^{S}x}\phi_{m'}^{S}+a'_{51}\chi_{3}^{S_{R}}e^{-iq_{h}^{S}x}\phi_{m'+1}^{S}+a'_{52}\chi_{4}^{S_{R}}e^{-iq_{h}^{S}x}\phi_{m'}^{S}+b'_{51}\chi_{1}^{S_{R}}e^{iq_{e}^{S}x}\phi_{m'}^{S}+b'_{52}\chi_{2}^{S_{R}}e^{iq_{e}^{S}x}\phi_{m'+1}^{S}, \hspace{0.7cm}\text{$x>a/2$}.
\end{cases}\\
{}\varphi_{6}(x)=\begin{cases}
k'_{61}\chi_{1}^{S}e^{-iq_{e}^{S}x}\phi_{m'-1}^{S}+k'_{62}\chi_{2}^{S}e^{-iq_{e}^{S}x}\phi_{m'}^{S}+l'_{61}\chi_{3}^{S}e^{iq_{h}^{S}x}\phi_{m'}^{S}+l'_{62}\chi_{4}^{S}e^{iq_{h}^{S}x}\phi_{m'-1}^{S}, \hspace{3.5cm}\text{$x<-a/2$},\\
c'_{61}\chi_{1}^{N}e^{iq_{e}(x+a/2)}\phi_{m'-1}^{S}+c'_{62}\chi_{2}^{N}e^{iq_{e}(x+a/2)}\phi_{m'}^{S}+d'_{61}\chi_{1}^{N}e^{-iq_{e}x}\phi_{m'-1}^{S}+d'_{62}\chi_{2}^{N}e^{-iq_{e}x}\phi_{m'}^{S}+e'_{61}\chi_{3}^{N}e^{iq_{h}x}\phi_{m'}^{S}+e'_{62}\chi_{4}^{N}e^{iq_{h}x}\phi_{m'-1}^{S}\\
+f'_{61}\chi_{3}^{N}e^{-iq_{h}(x+a/2)}\phi_{m'}^{S}+f'_{62}\chi_{4}^{N}e^{-iq_{h}(x+a/2)}\phi_{m'-1}^{S},\hspace{6.8cm}\text{$-a/2<x<0$},\\
g'_{61}\chi_{1}^{N}e^{-iq_{e}(x-a/2)}\phi_{m'-1}^{S}+g'_{62}\chi_{2}^{N}e^{-iq_{e}(x-a/2)}\phi_{m'}^{S}+h'_{61}\chi_{1}^{N}e^{iq_{e}x}\phi_{m'-1}^{S}+h'_{62}\chi_{2}^{N}e^{iq_{e}x}\phi_{m'}^{S}+i'_{61}\chi_{3}^{N}e^{-iq_{h}x}\phi_{m'}^{S}+i'_{62}\chi_{4}^{N}e^{-iq_{h}x}\phi_{m'-1}^{S}\\
+j'_{61}\chi_{3}^{N}e^{iq_{h}(x-a/2)}\phi_{m'}^{S}+j'_{62}\chi_{4}^{N}e^{iq_{h}(x-a/2)}\phi_{m'-1}^{S},\hspace{7.5cm}\text{$0<x<a/2$},\\
\chi_{2}^{S_{R}}e^{-iq_{e}^{S}x}\phi_{m'}^{S}+a'_{61}\chi_{3}^{S_{R}}e^{-iq_{h}^{S}x}\phi_{m'}^{S}+a'_{62}\chi_{4}^{S_{R}}e^{-iq_{h}^{S}x}\phi_{m'-1}^{S}+b'_{61}\chi_{1}^{S_{R}}e^{iq_{e}^{S}x}\phi_{m'-1}^{S}+b'_{62}\chi_{2}^{S_{R}}e^{iq_{e}^{S}x}\phi_{m'}^{S}, \hspace{0.7cm}\text{$x>a/2$}.
\end{cases}\\
\end{split}\nonumber
\end{eqnarray}
\begin{eqnarray}
\begin{split}
{}\varphi_{7}(x)=\begin{cases}
k'_{71}\chi_{1}^{S}e^{-iq_{e}^{S}x}\phi_{m'-1}^{S}+k'_{72}\chi_{2}^{S}e^{-iq_{e}^{S}x}\phi_{m'}^{S}+l'_{71}\chi_{3}^{S}e^{iq_{h}^{S}x}\phi_{m'}^{S}+l'_{72}\chi_{4}^{S}e^{iq_{h}^{S}x}\phi_{m'-1}^{S}, \hspace{3.5cm}\text{$x<-a/2$},\\
c'_{71}\chi_{1}^{N}e^{iq_{e}(x+a/2)}\phi_{m'-1}^{S}+c'_{72}\chi_{2}^{N}e^{iq_{e}(x+a/2)}\phi_{m'}^{S}+d'_{71}\chi_{1}^{N}e^{-iq_{e}x}\phi_{m'-1}^{S}+d'_{72}\chi_{2}^{N}e^{-iq_{e}x}\phi_{m'}^{S}+e'_{71}\chi_{3}^{N}e^{iq_{h}x}\phi_{m'}^{S}+e'_{72}\chi_{4}^{N}e^{iq_{h}x}\phi_{m'-1}^{S}\\
+f'_{71}\chi_{3}^{N}e^{-iq_{h}(x+a/2)}\phi_{m'}^{S}+f'_{72}\chi_{4}^{N}e^{-iq_{h}(x+a/2)}\phi_{m'-1}^{S},\hspace{6.8cm}\text{$-a/2<x<0$},\\
g'_{71}\chi_{1}^{N}e^{-iq_{e}(x-a/2)}\phi_{m'-1}^{S}+g'_{72}\chi_{2}^{N}e^{-iq_{e}(x-a/2)}\phi_{m'}^{S}+h'_{71}\chi_{1}^{N}e^{iq_{e}x}\phi_{m'-1}^{S}+h'_{72}\chi_{2}^{N}e^{iq_{e}x}\phi_{m'}^{S}+i'_{71}\chi_{3}^{N}e^{-iq_{h}x}\phi_{m'}^{S}+i'_{72}\chi_{4}^{N}e^{-iq_{h}x}\phi_{m'-1}^{S}\\
+j'_{71}\chi_{3}^{N}e^{iq_{h}(x-a/2)}\phi_{m'}^{S}+j'_{72}\chi_{4}^{N}e^{iq_{h}(x-a/2)}\phi_{m'-1}^{S},\hspace{7.5cm}\text{$0<x<a/2$},\\
\chi_{3}^{S_{R}}e^{iq_{h}^{S}x}\phi_{m'}^{S}+b'_{71}\chi_{3}^{S_{R}}e^{-iq_{h}^{S}x}\phi_{m'}^{S}+b'_{72}\chi_{4}^{S_{R}}e^{-ik_{h}^{S}x}\phi_{m'-1}^{S}+a'_{71}\chi_{1}^{S_{R}}e^{iq_{e}^{S}x}\phi_{m'-1}^{S}+a'_{72}\chi_{2}^{S_{R}}e^{iq_{e}^{S}x}\phi_{m'}^{S}, \hspace{0.7cm}\text{$x>a/2$}.
\end{cases}\\
{}\varphi_{8}(x)=\begin{cases}
k'_{81}\chi_{1}^{S}e^{-iq_{e}^{S}x}\phi_{m'}^{S}+k'_{82}\chi_{2}^{S}e^{-iq_{e}^{S}x}\phi_{m'+1}^{S}+l'_{81}\chi_{3}^{S}e^{iq_{h}^{S}x}\phi_{m'+1}^{S}+l'_{82}\chi_{4}^{S}e^{iq_{h}^{S}x}\phi_{m'}^{S}, \hspace{3.5cm}\text{$x<-a/2$},\\
c'_{81}\chi_{1}^{N}e^{iq_{e}(x+a/2)}\phi_{m'}^{S}+c'_{82}\chi_{2}^{N}e^{iq_{e}(x+a/2)}\phi_{m'+1}^{S}+d'_{81}\chi_{1}^{N}e^{-iq_{e}x}\phi_{m'}^{S}+d'_{82}\chi_{2}^{N}e^{-iq_{e}x}\phi_{m'+1}^{S}+e'_{81}\chi_{3}^{N}e^{iq_{h}x}\phi_{m'+1}^{S}+e'_{82}\chi_{4}^{N}e^{iq_{h}x}\phi_{m'}^{S}\\
+f'_{81}\chi_{3}^{N}e^{-iq_{h}(x+a/2)}\phi_{m'+1}^{S}+f'_{82}\chi_{4}^{N}e^{-iq_{h}(x+a/2)}\phi_{m'}^{S},\hspace{6.8cm}\text{$-a/2<x<0$},\\
g'_{81}\chi_{1}^{N}e^{-iq_{e}(x-a/2)}\phi_{m'}^{S}+g'_{82}\chi_{2}^{N}e^{-iq_{e}(x-a/2)}\phi_{m'+1}^{S}+h'_{81}\chi_{1}^{N}e^{iq_{e}x}\phi_{m'}^{S}+h'_{82}\chi_{2}^{N}e^{iq_{e}x}\phi_{m'+1}^{S}+i'_{81}\chi_{3}^{N}e^{-iq_{h}x}\phi_{m'+1}^{S}+i'_{82}\chi_{4}^{N}e^{-iq_{h}x}\phi_{m'}^{S},\\
+j'_{81}\chi_{3}^{N}e^{iq_{h}(x-a/2)}\phi_{m'+1}^{S}+j'_{82}\chi_{4}^{N}e^{iq_{h}(x-a/2)}\phi_{m'}^{S},\hspace{7.5cm}\text{$0<x<a/2$},\\
\chi_{4}^{S_{R}}e^{iq_{h}^{S}x}\phi_{m'}^{S}+b'_{81}\chi_{3}^{S_{R}}e^{-iq_{h}^{S}x}\phi_{m'+1}^{S}+b'_{82}\chi_{4}^{S_{R}}e^{-iq_{h}^{S}x}\phi_{m'}^{S}+a'_{81}\chi_{1}^{S_{R}}e^{iq_{e}^{S}x}\phi_{m'}^{S}+a'_{82}\chi_{2}^{S_{R}}e^{iq_{e}^{S}x}\phi_{m'+1}^{S}, \hspace{0.7cm}\text{$x>a/2$}.
\end{cases}\\
\end{split}
\label{wavv}
\end{eqnarray}}
\normalsize
where $\chi_{1}^{S_{R}}=\begin{pmatrix}
ue^{i\varphi}\\
0\\
0\\
v
\end{pmatrix}$, $\chi_{2}^{S_{R}}=\begin{pmatrix}
0\\
-ue^{i\varphi}\\
v\\
0
\end{pmatrix}$, $\chi_{3}^{S_{R}}=\begin{pmatrix}
0\\
-ve^{i\varphi}\\
u\\
0
\end{pmatrix}$ and, $\chi_{4}^{S_{R}}=\begin{pmatrix}
ve^{i\varphi}\\
0\\
0\\
u
\end{pmatrix}$. In Eq.~\eqref{wavv}, $\varphi_{1}$, $\varphi_{2}$, $\varphi_{3}$, and $\varphi_{4}$ represent the wavefunctions when spin up electronlike quasiparticle, spin down electronlike quasiparticle, spin up holelike quasiparticle and spin down holelike quasiparticle are injected from left superconducting region respectively, while $\varphi_{5}$, $\varphi_{6}$, $\varphi_{7}$, and $\varphi_{8}$ represent the wavefunctions when spin up electronlike quasiparticle, spin down electronlike quasiparticle, spin up holelike quasiparticle and spin down holelike quasiparticle are injected from right superconducting region respectively. The Andreev and normal reflection coefficients are $a'_{mn}$ and $b'_{mn}$ respectively, while the transmission coefficients for electronlike and holelike quasiparticles are $k'_{mn}$ and $l'_{mn}$ respectively. If we diagonalize the Hamiltonian $(H_{BdG}^{\mbox{S-N$_{1}$-sf-N$_{2}$-S}})^{*}(-k)$ instead of $H_{BdG}^{\mbox{S-N$_{1}$-sf-N$_{2}$-S}}(k)$ we will get conjugated processes $\tilde{\varphi_{i}}$ which is necessary to form the retarded Green's functions in next subsection. For our setup we note that $\tilde{\chi_{i}}^{N(S)}=\chi_{i}^{N(S)}$ and $\tilde{\chi_{1}}^{S_{R}}=\begin{pmatrix}
ue^{-i\varphi}\\
0\\
0\\
v
\end{pmatrix}$, $\tilde{\chi_{2}}^{S_{R}}=\begin{pmatrix}
0\\
-ue^{-i\varphi}\\
v\\
0
\end{pmatrix}$, $\tilde{\chi_{3}}^{S_{R}}=\begin{pmatrix}
0\\
-ve^{-i\varphi}\\
u\\
0
\end{pmatrix}$ and, $\tilde{\chi_{4}}^{S_{R}}=\begin{pmatrix}
ve^{-i\varphi}\\
0\\
0\\
u
\end{pmatrix}$.}
{\section{Effect of interface transparency on superconducting pairing}
In this section, we discuss the effect of interface transparency on Even-SS and Odd-EST pairings. In Fig.~12, we plot the induced Even-SS and Odd-EST pairings as function of interface transparency $Z$ for (a) N$_{1}$-SF-N$_{2}$-S junction and for (b) S-N$_{1}$-SF-N$_{2}$-S junction. We see that the magnitudes of Even-SS and Odd-EST pairings decrease with increasing $Z$. However, in the tunneling limit ($Z\gg0$), Odd-EST pairing vanishes, while Even-SS pairing is still finite with small magnitude. Thus, $p$-wave pairing is more fragile to interface transparency than $s$-wave pairing. However, Odd-EST pairing increases and is much larger than Even-SS pairing at intermediate $Z$ for N$_{1}$-SF-N$_{2}$-S junction and at low $Z$ for S-N$_{1}$-SF-N$_{2}$-S junction.
\begin{figure}[h]
\centering{\includegraphics[width=.8\textwidth]{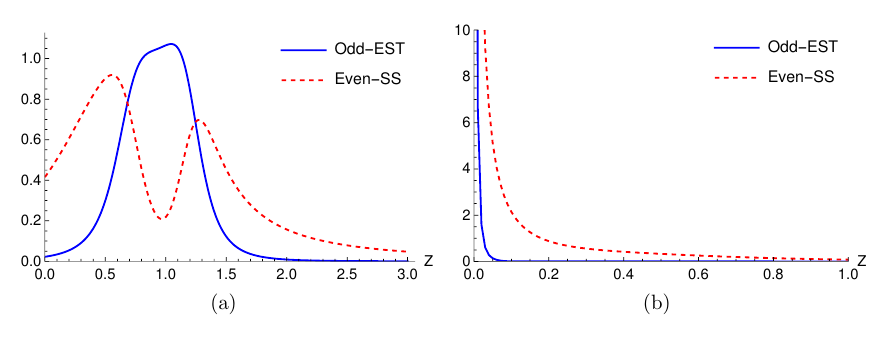}}
\caption{\small \sl The magnitudes of the induced Even-SS and Odd-EST pairing vs. interface transparency $Z$ for (a) N$_{1}$-SF-N$_{2}$-S junction and for (b) S-N$_{1}$-SF-N$_{2}$-S junction. Parameters: $S=\frac{1}{2}$, $\mathcal{J}=4.5$ (Fig.~12(a)), $\mathcal{J}=2.27$ (Fig.~12(b)) $x=0$, $x'=0$, $E_{F}=30\Delta$, $k_{F}\xi=2$, $k_{F}a=0.85\pi$ (Fig.~12(a)), $k_{F}a=\pi/8$ (Fig.~12(b)), $\varphi=\pi/8$, {$\omega\rightarrow0$}.}
\end{figure}
\section{Odd-frequency superconducting pairing: Effect of Finite temperature}
In the main text, we have discussed the results for SS and ST pairings at zero temperature. In this section SS and ST pairings are studied at finite temperature. To compute superconducting pairing at finite temperature, anomalous Green's function is written in Matsubara representation as\cite{feng}-
\begin{equation}
\label{greenFT}
\sum_{\omega_{n}>0} G^{r}_{eh}(x,x',i\omega_{n})=i\sum_{\mu=0}^{3}f_{\mu}^{r}\sigma_{\mu}\sigma_{2},
\end{equation}
where $\omega_{n}=\frac{\pi}{\beta}(2n+1)$ are Matsubara frequencies and $n=0,\pm1,\pm2,\pm3,....$
\begin{figure}[h]
\centering{\includegraphics[width=.8\textwidth]{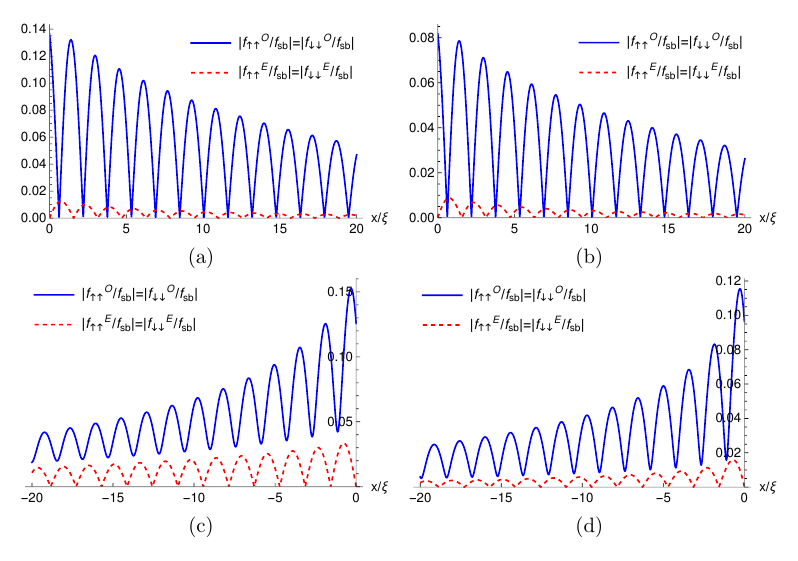}}
\caption{\small \sl The magnitudes of the induced Even-EST and Odd-EST pairing vs. position $x$ at finite temperature for (a) and (b) metal-spin flipper-metal-Superconductor (N$_{1}$-SF-N$_{2}$-S) junction, for (c) and (d) Superconductor-metal-spin flipper-metal-Superconductor (S-N$_{1}$-SF-N$_{2}$-S) junction. Parameters: $S=\frac{1}{2}$, $Z=0.776$ (for (a) and (b)), $Z=0$ (for (c) and (d)), $\mathcal{J}=4.5$ (for (a) and (b)), $\mathcal{J}=2.27$ (for (c) and (d)),  $x'=0$, $E_{F}=30\Delta$, $k_{F}\xi=2$, $k_{F}a=0.85\pi$ (for (a) and (b)), $k_{F}a=\pi/8$ (for (c) and (d)), $\varphi=\pi/8$,  $T=0.05T_{c}$ (for (a) and (c)), $T=0.25T_{c}$ (for (b) and (d)). In figure we normalize the magnitudes of the induced superconducting pairing with respect to the magnitude of SS pairing ($|f_{sb}|$) in the bulk superconductors.}
\end{figure}
\begin{figure}[h]
\centering{\includegraphics[width=.8\textwidth]{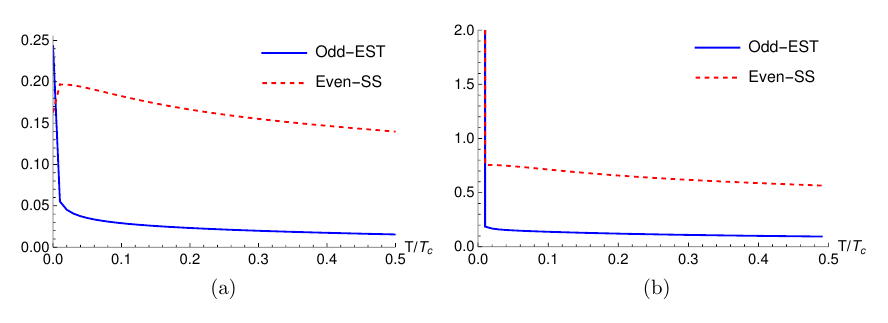}}
\caption{\small \sl The magnitudes of the Even-SS and Odd-EST pairing vs. temperature for (a) metal-spin flipper-metal-Superconductor (N$_{1}$-SF-N$_{2}$-S) junction and for (b) Superconductor-metal-spin flipper-metal-Superconductor (S-N$_{1}$-SF-N$_{2}$-S) junction. Parameters: $S=\frac{1}{2}$, $Z=0.776$ (for (a)), $Z=0$ (for (b)), $\mathcal{J}=4.5$ (for (a)), $\mathcal{J}=2.27$ (for (b)), $x=0.75\xi$ (for (a)), $x=-0.25\xi$ (for (b)), $x'=0$, $E_{F}=30\Delta$, $k_{F}\xi=2$, $k_{F}a=0.85\pi$ (for (a)), $k_{F}a=\pi/8$ (for (b)), $\varphi=\pi/8$. In figure we normalize the magnitudes of the induced superconducting pairing ($|f_{sb}|$) with respect to the magnitude of SS pairing in the bulk superconductors.}
\end{figure}
In Eq.~\eqref{greenFT}, we take summation only $\omega_{n}>0$ since all pairings become odd functions of frequency. From Eq.~\eqref{greenFT}, we can calculate Even-/Odd-SS and Even-/Odd-ST pairings. In Fig.~13 we plot Even-EST and Odd-EST pairing at two different finite temperatures $T=0.05T_{c}$ (13~(a),(c)) and $T=0.25T_{c}$ (13~(b),(d)) for N$_{1}$-SF-N$_{2}$-S junction (13~(a),(b)) and S-N$_{1}$-SF-N$_{2}$-S junction (13~(c),(d)). We notice that Odd-EST pairing is still much larger than Even-EST pairing. This behavior is seen at any finite temperature below critical temperature $T_{c}$. In Fig.~14, Odd-EST pairing and Even-SS pairing are plotted as function of temperature in the superconducting region for N$_{1}$-SF-N$_{2}$-S junction (14~(a)) and S-N$_{1}$-SF-N$_{2}$-S junction (14~(b)). We see that at low temperatures, Odd-EST pairing is much larger than Even-SS pairing, however at high temperatures Even-SS pairing is larger than Odd-EST pairing. In Figs.~13, 14, we normalize the magnitudes of the induced superconducting pairing with respect to the magnitude of SS pairing in the bulk superconductors,
\begin{equation}
f_{sb}=\sum_{\omega_{n}}\frac{\Delta}{\sqrt{\omega_{n}^2+\Delta^2}}.
\end{equation}
The superconducting gap parameter $\Delta$ depends on temperature via $\Delta(T)=\Delta(0)\tanh(1.74\sqrt{T_{c}/T-1})$, where $T_{c}$ is the critical temperature\cite{annu}.}

\section{Explicit form of expressions for Anomalous Green's functions}
In this section, we give an explicit form of expressions for anomalous Green's functions ($G^{r}_{eh}$). We compute SS and ST pairing using $G^{r}_{eh}$.
\subsection{Metal (N$_{1}$)-Spin flipper (sf)-Metal ( N$_{2}$)-Superconductor (S) junction}
In case of N$_{1}$-sf-N$_{2}$-S junction, $G^{r}$ are determined by putting wavefunctions from Eq.~\eqref{wav} into Eq.~\eqref{RGF} with $\it{a}_{ij}$ and $\it{b}_{ij}$ found from Eqs.~\eqref{bc1}-\eqref{bc4}. For $G^{r}_{eh}$ we get,
{
\begin{equation}
\label{supamp}
\begin{split}
[G^{r}_{eh}]_{\uparrow\uparrow}&=-\frac{\eta}{2i(u^2-v^2)}\Bigg[\frac{\it{b}_{61}e^{iq_{e}^{S}(x+x')}uv+\it{a}_{62}e^{i(q_{e}^{S}x'-q_{h}^{S}x)}v^2}{q_{e}^{S}}+\frac{\it{a}_{62}e^{i(q_{e}^{S}x-q_{h}^{S}x')}u^2-\it{b}_{72}e^{-iq_{h}^{S}(x+x')}uv}{q_{h}^{S}}\Bigg],\\
[G^{r}_{eh}]_{\downarrow\downarrow}&=-\frac{\eta}{2i(u^2-v^2)}\Bigg[\frac{-\it{b}_{52}e^{iq_{e}^{S}(x+x')}uv+\it{a}_{51}e^{i(q_{e}^{S}x'-q_{h}^{S}x)}v^2}{q_{e}^{S}}+\frac{\it{a}_{51}e^{i(q_{e}^{S}x-q_{h}^{S}x')}u^2+\it{b}_{81}e^{-iq_{h}^{S}(x+x')}uv}{q_{h}^{S}}\Bigg],\\
[G^{r}_{eh}]_{\uparrow\downarrow}&=\frac{\eta}{2i(u^2-v^2)}\Bigg[\frac{e^{iq_{e}^{S}|x-x'|}uv+\it{b}_{51}e^{iq_{e}^{S}(x+x')}uv+\it{a}_{81}e^{i(q_{e}^{S}x'-q_{h}^{S}x)}v^2}{q_{e}^{S}}\\&+\frac{\it{a}_{81}e^{i(q_{e}^{S}x-q_{h}^{S}x')}u^2+\it{b}_{82}e^{-iq_{h}^{S}(x+x')}uv+e^{-iq_{h}^{S}|x-x'|}uv}{q_{h}^{S}}\Bigg],\\
[G^{r}_{eh}]_{\downarrow\uparrow}&=-\frac{\eta}{2i(u^2-v^2)}\Bigg[\frac{e^{iq_{e}^{S}|x-x'|}uv+\it{b}_{62}e^{iq_{e}^{S}(x+x')}uv-\it{a}_{72}e^{i(q_{e}^{S}x'-q_{h}^{S}x)}v^2}{q_{e}^{S}}\\&+\frac{-\it{a}_{72}e^{i(q_{e}^{S}x-q_{h}^{S}x')}u^2+\it{b}_{71}e^{-iq_{h}^{S}(x+x')}uv+e^{-iq_{h}^{S}|x-x'|}uv}{q_{h}^{S}}\Bigg]
\end{split}
\end{equation}}
\subsection{Superconductor (S)-Metal (N$_{1}$)-Spin flipper (sf)-Metal (N$_{2}$)-Superconductor (S) junction}
In case of S-N$_{1}$-sf-N$_{2}$-S junction, $G^{r}$ are obtained by putting wavefunctions from Eq.~\eqref{wavv} into Eq.~\eqref{RGF} with $a'_{ij}$ and $b'_{ij}$ found from Eqs.~\eqref{bc5}-\eqref{bc55}. For $G^{r}_{eh}$ we get
{
\begin{equation}
\label{suppamp}
\begin{split}
[G^{r}_{eh}]_{\uparrow\uparrow}&=-\frac{\eta}{2i(u^2-v^2)}\Bigg[\frac{b'_{21}e^{-iq_{e}^{S}(x+x')}uv+a'_{22}e^{-i(q_{e}^{S}x'-q_{h}^{S}x)}v^2}{q_{e}^{S}}-\frac{a'_{31}e^{-i(q_{e}^{S}x-q_{h}^{S}x')}u^2+b'_{32}e^{iq_{h}^{S}(x+x')}uv}{q_{h}^{S}}\Bigg],\\
[G^{r}_{eh}]_{\downarrow\downarrow}&=-\frac{\eta}{2i(u^2-v^2)}\Bigg[\frac{-b'_{12}e^{-iq_{e}^{S}(x+x')}uv+a'_{11}e^{-i(q_{e}^{S}x'-q_{h}^{S}x)}v^2}{q_{e}^{S}}-\frac{a'_{42}e^{-i(q_{e}^{S}x-q_{h}^{S}x')}u^2-b'_{41}e^{iq_{h}^{S}(x+x')}uv}{q_{h}^{S}}\Bigg],\\
[G^{r}_{eh}]_{\uparrow\downarrow}&=\frac{\eta}{2i(u^2-v^2)}\Bigg[\frac{e^{iq_{e}^{S}|x-x'|}uv+b'_{11}e^{-iq_{e}^{S}(x+x')}uv+a'_{12}e^{-i(q_{e}^{S}x'-q_{h}^{S}x)}v^2}{q_{e}^{S}}\\&+\frac{a'_{41}e^{-i(q_{e}^{S}x-q_{h}^{S}x')}u^2+b'_{42}e^{iq_{h}^{S}(x+x')}uv+e^{-iq_{h}^{S}|x-x'|}uv}{q_{h}^{S}}\Bigg],\\
[G^{r}_{eh}]_{\downarrow\uparrow}&=-\frac{\eta}{2i(u^2-v^2)}\Bigg[\frac{e^{iq_{e}^{S}|x-x'|}uv+b'_{22}e^{-iq_{e}^{S}(x+x')}uv-a'_{21}e^{-i(q_{e}^{S}x'-q_{h}^{S}x)}v^2}{q_{e}^{S}}\\&+\frac{-a'_{32}e^{-i(q_{e}^{S}x-q_{h}^{S}x')}u^2+b'_{31}e^{iq_{h}^{S}(x+x')}uv+e^{-iq_{h}^{S}|x-x'|}uv}{q_{h}^{S}}\Bigg].
\end{split}
\end{equation}}

\end{document}